# Metal Halide Perovskite/Chalcohalide Heterojunctions for the Photoinduced Oxidative Coupling of *p*-substituted Thiophenols

*Anna Cabona,[1,2,4] Stefano Toso,[1,3] Alejandro Cortés-Villena,[4] Ignacio Rosa-Pardo,[4] Mirko Prato,[5] Michele Ferri,[1] Julia Perez-Prieto,[4]* Ilka Kriegel,[2]* Liberato Manna[1]* and Raquel E. Galian[4]**

[1] Nanochemistry Department, Italian Institute of Technology, 16163 Genova, Italy;

[2] Department of Applied Science and Technology, Politecnico di Torino, 10129 Turin, Italy;

[3] Lund University, Division of Chemical Physics, Naturvetarvägen 14, 221 00 Lund, Sweden;

[4] Institute of Molecular Science, University of Valencia, c/Catedrático José Beltrán Martínez 2, 46980 Paterna, Valencia, Spain;

[5] Materials Characterization, Italian Institute of Technology, 16163 Genova, Italy;

**Corresponding Authors**

Julia Pérez-Prieto: julia.perez@uv.es, Ilka Kriegel: ilka.kriegel@polito.it, Liberato Manna: liberato.manna@iit.it, Raquel E. Galian: raquel.galian@uv.es

ABSTRACT

The introduction of a semiconductor-semiconductor junction is an effective strategy to enhance the photocatalytic performance of perovskite nanocrystal-based systems. Herein, we optimized the synthesis of $CsPbX_3/Pb_4S_3X_2$ (X= Cl, Br, I) perovskites-chalcohalides heterostructures, whose band alignment can be tuned by halide composition. As a *proof-of-concept*, we evaluated the photooxidative coupling of *p*-substituted thiophenols at room temperature, under visible-light, air, and without sacrificial electron donor. Notably, $CsPbBr_3/Pb_4S_3Br_2$ achieved up to 94 % selectivity toward disulfide (*p*-$OCH_3$ thiophenol with a turnover number of 14300) highlighting the crucial role of the type-II heterojunction to promote charge separation and efficient electron delocalization across the junction.

TEXT

Colloidal nanocrystal heterostructures (HSs) are nanoparticles composed of at least two distinct materials sharing an interface[1,2] exhibiting properties that can be distinct from their individual components. Semiconductor–semiconductor HSs are especially promising for light-harvesting applications,[3] as both domains can absorb light and generate electron–hole pairs. Proper band alignment at the heterojunction can enhance charge separation and reduce recombination, [4] making these architectures attractive for photocatalysis.[5]

HSs based on lead halide perovskites have emerged as promising photocatalysts,[6] as they combine tunable band gaps[7] and the intrinsic defect tolerance[8] of $CsPbX_3$ with the additional electronic tunability of the heterojunctions. Indeed, significant progress has been achieved using $CsPbBr_3$ nanocrystals as photocatalysts for organic transformations.[9] However, improving the photocatalytic performances of metal halide perovskites, beyond their intrinsic instability under thermal and humid conditions,[10] requires addressing several crucial factors such as their optical absorption properties, the generation and separation of photogenerated charge carriers, and the surface reaction kinetics. A highly promising strategies to enhance charge separation efficiency is coupling perovskites with a secondary semiconductor. This combination allows for an improved surface reaction rate through advanced configurations, such as Type II, Z-scheme, and S-scheme heterojunctions.[11] In this context, $CsPbX_3/Pb_4S_3Y_2$ (X, Y = Cl, Br, I) perovskite–chalcohalide epitaxial HSs, initially reported by some of the authors,[12] have recently emerged as promising photocatalysts, where strong interfacial coupling enhances charge separation.[12] For instance, Pradhan et al.[13] recently demonstrated that $CsPbBr_3/Pb_4S_3Br_2$ HSs outperform $CsPbBr_3$ nanocrystals (NCs) in $CO_2$ photoreduction due to suppressed radiative recombination and efficient charge transfer across the heterojunction.

Most perovskite-based heterostructures photocatalysts reported in the literature present a non-epitaxial interface between the two semiconductors,[14] where the benefits of the architecture mainly rely on the heterojunction type rather than on interface quality.[15] In contrast, the proposed perovskite/chalcohalide $CsPbX_3/Pb_4S_3Y_2$ HSs present several advantages over the reported HSs: a) an epitaxial interface, which enables a potentially defect-free junction, thereby reducing the probability of recombination centers; b) a type-II heterojunction, with efficient electron delocalization across the nearly isoenergetic conduction bands of both semiconductors; and c) a dimer morphology that could improve the number of effective contact points with the substrate.

Despite this potential, their application in organic transformations in non-polar media remains largely unexplored. To address this gap, we investigate the band alignment of three different $CsPbX_3/Pb_4S_3Y_2$ (X, Y = Cl, Br, I) HSs, as interfacial energy level matching is crucial to design an effective photocatalyst for specific target reactions.

Based on this knowledge, we selected the oxidative coupling of thiophenols into disulfides as a *proof-of-concept* reaction to directly assess how the use of a semiconductor-semiconductor heterostructure catalyst enhances the photocatalytic performance. This model transformation, relevant across (bio)chemistry[16] and industry,[17] enables direct evaluation of the advantages of semiconductor heterojunctions over single-phase nanocrystals. Compared to conventional methods, this photocatalytic approach offers a milder and more sustainable alternative, avoiding over-oxidation and simplifying purification.[18] To assess the photocatalytic activity of the $CsPbX_3/Pb_4S_3Y_2$ HSs in the oxidation of *p*-thiophenols we compared their reactivity against their individual components ($CsPbX_3$ and $Pb_4S_3X_2$ NCs). Among all HSs we tested, $CsPbBr_3/Pb_4S_3Br_2$ showed the highest yield (81%) and selectivity (87%), outperforming stand-alone NCs and suggesting efficient charge separation at the interface. These results were attained in only 90 minutes upon 450 nm illumination, a significantly shorter time than previous studies with perovskite-based photocatalyst.[19] These results underscore the pivotal role of heterojunction configuration in enhances charge separation and photocatalytic performances.

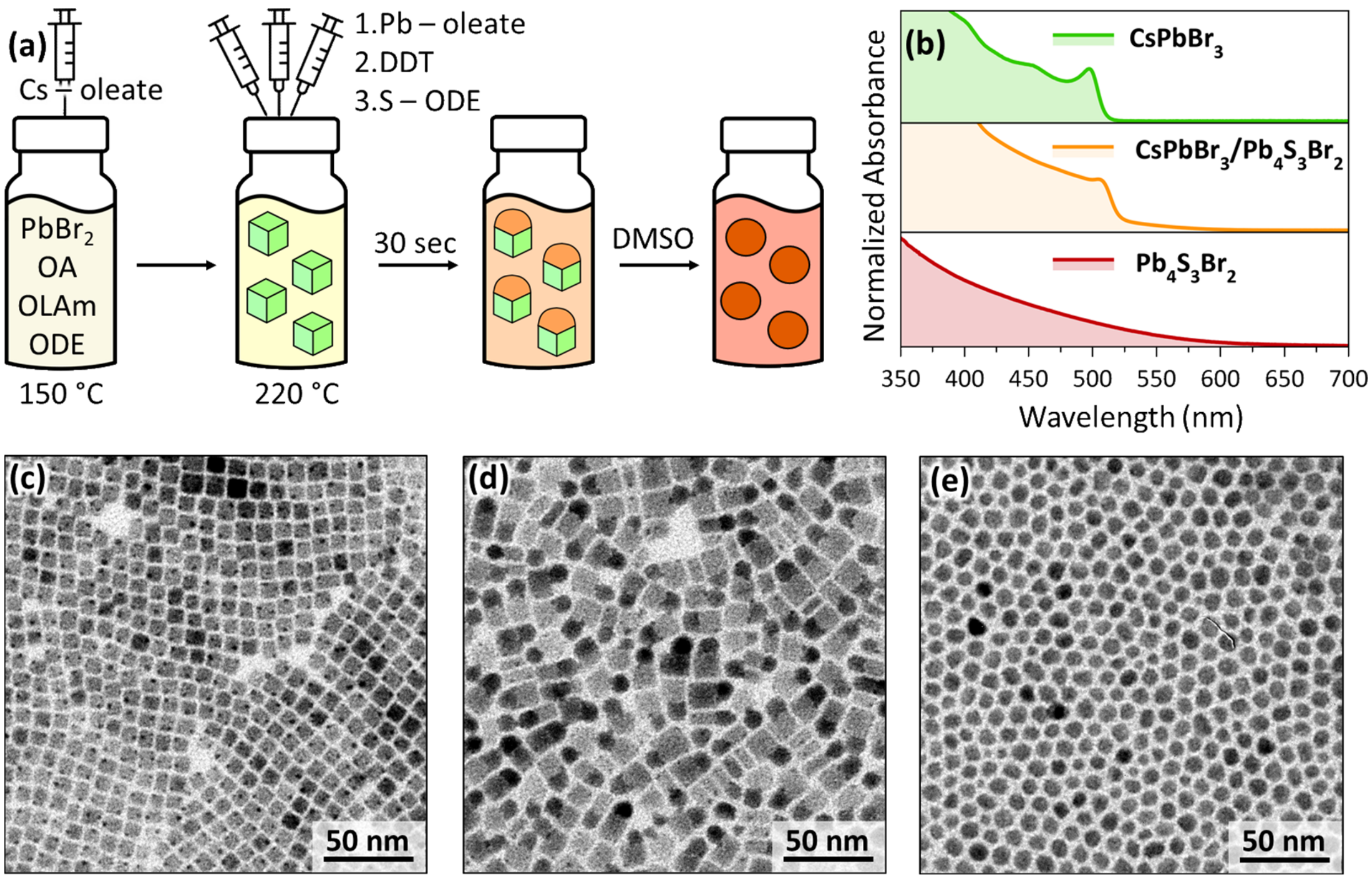

**Figure 1. Synthesis of $CsPbBr_3/Pb_4S_3Br_2$ HSs.** a) Scheme of the synthetic route used to obtain $CsPbBr_3$ NCs and $CsPbBr_3/Pb_4S_3Br_2$ HSs, followed by the selective etching of the perovskite domain to isolate free-standing $Pb_4S_3Br_2$ NCs. b) Absorption spectra of $CsPbBr_3$ NCs (green), $CsPbBr_3/Pb_4S_3Br_2$ HSs (orange) and $Pb_4S_3Br_2$ NCs (red). c-e) TEM images of $CsPbBr_3$ NCs (c), $CsPbBr_3/Pb_4S_3Br_2$ HSs (d), and $Pb_4S_3Br_2$ NCs (e).

The first synthetic protocol developed for $CsPbBr_3/Pb_4S_3Br_3$ HSs relied on perovskite nanoclusters as precursors, which were reacted with a sulfur source to obtain the product.[12] While producing well-defined heterostructures, this protocol has a low reaction yield, hampering extensive testing. Therefore, we here optimized an alternative published protocol,[13] using pre-formed $CsPbBr_3$ nanocrystals reacted with lead oleate, 1-dodecanthiol (DDT) and sulfur in 1-octadecene (S-ODE). This approach significantly increases the yield (130 mg/batch), albeit at the cost of a less controlled morphology and a higher fraction of stand-alone $CsPbBr_3$ nanocrystals (≈20-25 %, quantified by analyzing more than 500 NCs from TEM images). Figure 1a is a scheme of the protocol, where $CsPbX_3$ NCs synthesized *via* a hot-injection method[20] are immediately reacted, without purification, with the chalcohalide precursors injected at 220 °C. "Free standing" $CsPbX_3$ NCs (used later as reference) were obtained by interrupting the synthesis after the first step, while $Pb_4S_3Y_2$ nanocrystals were recovered by selective etching of the perovskite domain in the corresponding HSs using dimethyl sulfoxide (DMSO).[21] Figure 1b shows the typical optical absorption spectra of the HSs and isolated NCs for the X=Y=Br case, with characteristic absorption features of the perovskite highlighted in green and the broad and featureless absorption of $Pb_4S_3Br_2$ NCs in red. As expected, the HSs absorption spectrum combines both features of the free standing NCs. Representative TEM images of the three samples are displayed in Figure 1c-e (see Figures S1-S3 for additional characterization).

The presence of two independent semiconductor domains allows for some tuning of the band gap to target specific reactions, in principle expanding the scope of HSs as photocatalysts. For instance, replacing $PbBr_2$ with $PbCl_2$ in the reaction yields $CsPbCl_3/Pb_4S_3Cl_2$ HSs (Figure 2a-b) with only minor adjustments required to the synthesis (see methods and Figures S4-S7). Conversely, the direct synthesis of $CsPbI_3/Pb_4S_3I_2$ HSs was unsuccessful, probably due to the increased lattice mismatch of $CsPbI_3/Pb_4S_3I_2$ compared to the Br- and Cl-analogues.[12] However, $CsPbI_3$-based HSs can still be obtained *via* halide exchange from $CsPbBr_3/Pb_4S_3Br_2$ HSs, as shown by us in a previous work.[12] Due to the low ionic mobility in the chalcohalide domain, such reaction primarily affects the perovskite domain,[12] resulting in $CsPbI_3/Pb_4S_3Br_2$ HSs (Figure 2c-d, Figure S8). The exchange was tracked by the redshift of the perovskite absorption onset, while the contribution from the chalcohalides domain remained essentially unchanged. (Figure 2c).

The combination of these strategies allows $CsPbX_3/Pb_4S_3X_2$ HSs to cover the whole visible spectrum (Figure 2e). The band alignment and redox properties of “free-standing” $CsPbX_3$ and $Pb_4S_3X_2$ NCs were characterized by combining optical absorption spectroscopy, to determine the optical band gap, with ultraviolet photoelectron spectroscopy (UPS) and ambient pressure photoemission spectroscopy (APS), to establish the position of the valence band (VB) edge with respect to the vacuum level (Figures S9-S13). While literature is rich of reports on the band positions of $CsPbX_3$,[22] very limited experimental data are available for $Pb_4S_3X_2$ chalcohalides.[21,23] Our results confirm the current computational predictions,[23,12] highlighting that the band gap of lead chalcohalides is only minimally affected by the halide composition, likely due to the band edge states being predominantly derived from $Pb^{2+}$ (conduction band) and $S^{2-}$ (valence band) orbitals.[23] We also confirmed the type-I and type-II band alignments predicted respectively for $CsPbCl_3/Pb_4S_3Cl_2$ and $CsPbBr_3/Pb_4S_3Br_2$ HSs, which explains the photoluminescence quenching observed in both systems. Notably, for the $CsPbBr_3/Pb_4S_3Br_2$ HS the conduction band edges of the two materials are closely aligned in energy (–3.4 eV for $CsPbBr_3$ and –3.3 eV for $Pb_4S_3Br_2$), thus facilitating electron delocalization across the HS.

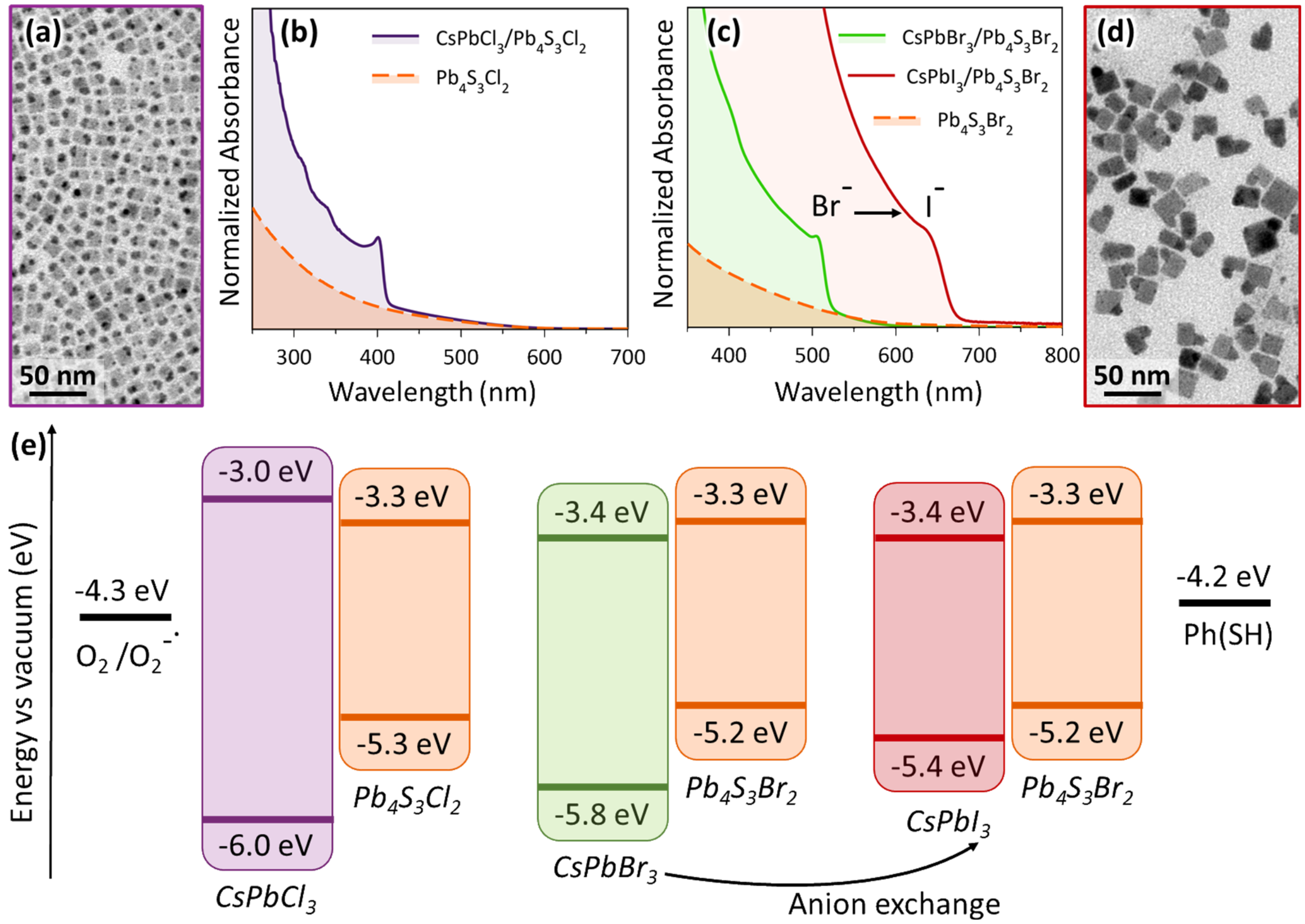

**Figure 2. Synthesis of Cl-/I-based HSs and band alignments.** a) TEM images of $CsPbCl_3/Pb_4S_3Cl_2$ HSs. b) Absorption spectra of $CsPbCl_3/Pb_4S_3Cl_2$ HSs (purple line) and $Pb_4S_3Cl_2$ chalcohalides (orange line). c) Absorption spectra of $CsPbBr_3/Pb_4S_3Br_2$ HSs (green line), $CsPbI_3/Pb_4S_3Br_2$ HSs after anion exchange (red line) and $Pb_4S_3Br_2$ chalcohalides NCs (orange line). d) TEM images of $CsPbI_3/Pb_4S_3Br_2$ HSs after anion exchange. (e) Band alignments of the Valence Band (VB) and Conduction Band (CB) of different HSs, compared with the highest-occupied molecular orbital (HOMO) of thiophenol (PhSH) and the reduction potential of $O_2$ to superoxide anion.

According to the energy levels estimated for the $CsPbX_3/Pb_4S_3X_2$ and the HOMO value for thiophenol (–4.2 eV, calculated from cyclovoltammetry, Figure S14), the $CsPbBr_3/Pb_4S_3Br_2$ HSs present optimal band levels to perform the photocatalytic reaction. In the case of $CsPbCl_3/Pb_4S_3Cl_2$

HSs, the type-I alignment would lead to a competitive non-radiative recombination in the chalcohalide domain, limiting charge separation, while $CsPbI_3/Pb_4S_3Br_2$ HSs requires an additional halide exchange step in the synthesis without offering sensible advantage over the $CsPbBr_3/Pb_4S_3Br_2$ HSs.

Based on these considerations, we selected as a benchmark the photooxidation of PhSH (**1a**) to disulfide (**1b**) catalyzed by $CsPbBr_3/Pb_4S_3Br_2$ HSs under 450 nm LED excitation (Scheme 1) and without an external sacrificial electron donor, as discussed in detail later (see methods, Figure S15-16 and Table S1).

**Scheme 1.** Standard conditions for the photooxidative coupling of *p*-substituted thiophenols.

2 X

NCs (1 mg/mL)
hv (450/405 nm), RT, 90 min
Cyclohexane
Air atmosphere

R = -H (1a)/ -$OCH_3$ (2a)/ -Br (3a)

R = -H (1b)/ -$OCH_3$ (2b)/ -Br (3b)

Preliminary solvent screening showed that cyclohexane gave the highest yield (81%, Table 1, Figure S17), outperforming hexane likely due to structural features that enhance solvation of both nanoparticles and substrate, improving photocatalyst–substrate interactions.[24] A 1:1 $CH_2Cl_2$/cyclohexane mixture gave high yield (76%) but lower selectivity, which is unfavorable for further photocatalytic investigations. In contrast, the photoreaction carried out in $CH_2Cl_2$ yielded even lower yield (33%). This was attributed to the partial anion exchange between $Br^-$ in the perovskite domain and $Cl^-$ ions from $CH_2Cl_2$ as a result of the C-Cl bound cleavage, as previously reported by Wu et al. (Figure S18).[19] These results highlight that reaction kinetics depend also on substrate diffusion and solvation energy, with the solvent influencing substrate adsorption on the NC surface and consequently the S–S bond formation.

**Table 1**. Standard conditions[a] using different solvents for the photooxidative coupling of the thiophenol.

| Entry | Solvent | Conversion (%) | 1b Yield (%) | Selectivity (%) |
|---|---|---|---|---|
| **1** | Cyclohexane | 93.0 | 81 | 87 |
| **2** | Cyclohexane: $CH_2Cl_2$ (1:1) | 98.0 | 76 | 76 |
| **3** | Hexane | 88.0 | 57 | 65 |
| **4** | $CH_2Cl_2$ | 100.0 | 33 | 33 |

[a] Conditions: thiophenol (0.034 mmol), $CsPbBr_3/Pb_4S_3Br_2$ HSs (1 mg) in 1 mL of solvent after 90 min of irradiation ($\lambda_{max}$ = 450 nm) at 20 °C under air. Yields are determined by GC-MS using biphenyl as internal standard.

The photocatalytic activity of the HSs was compared with single-component $CsPbBr_3$ and $Pb_4S_3Br_2$ NCs, using 1 mg of photoactive material (estimated by TGA analysis Figure S19). Although HSs and single-component NCs photocatalyzed the reaction (Table 2), both product yield and selectivity were significantly improved for $CsPbBr_3/Pb_4S_3Br_2$ HSs, and this can be ascribed to the effective charge separation induced by the electronic junction in the HSs. Control experiments conducted in the absence of the photocatalyst or light resulted in only 2 % or 10 % of **1b**, respectively, further confirming the photocatalytic activity of the HSs. The turnover number (TON) and turnover frequency (TOF) were determined according to the method previously reported by some of the authors.[25] The HSs exhibited the highest photocatalytic performance, with a TON of 14300 and a TOF of 9560, while significantly lower values were obtained for the $CsPbBr_3$ (TON = 3680; TOF = 2460) and $Pb_4S_3Br_2$ nanocrystals (TON = 2070; TOF = 1380).

Additionally, the $CsPbBr_3/Pb_4S_3Br_2$ HSs outperform other photocatalysts (Table S2) in terms of reduced reaction time, with high TON and TOF numbers.

**Table 2.** Standard conditions using different photocatalysts (bromide based) for the coupling of the thiophenol and *p*-substituted thiophenols.

| Entry | Variance | R substituent | Photocatalyst | Conversion (%) | 1b Yield (%) | Selectivity (%) |
|---|---|---|---|---|---|---|
| **1** | None | -H | $CsPbBr_3$ $/Pb_4S_3Br_2$ | 93.0 | 81 ± 1* | 87 |
| **2** | None | -H | $CsPbBr_3$ | 73.4 | 50 ± 4* | 72 |
| **3** | None | -H | $Pb_4S_3Br_2$ | 71.6 | 59 ± 4* | 86 |
| **4** | Without photocatalyst | -H | - | 19 | 2 | - |
| **5** | Without light | -H | $CsPbBr_3$ $/Pb_4S_3Br_2$ | 51 | 10 | - |
| **6** | Without light | -H | $CsPbBr_3$ | 45 | 4 | - |
| **7** | Without light | -H | $Pb_4S_3Br_2$ | 40 | 7 | - |
| **8** | Under nitrogen | -H | $CsPbBr_3$ $/Pb_4S_3Br_2$ | 61 | 22 | - |
| **9** | In nitrogen | -H | $CsPbBr_3$ | 40 | 2 | - |
| **10** | In nitrogen | -H | $Pb_4S_3Br_2$ | 40 | 5 | - |
| **11** | None | $-OCH_3$ | $CsPbBr_3$ $/Pb_4S_3Br_2$ | 100 | 94 | 94 |
| **12** | None | $-OCH_3$ | $CsPbBr_3$ | 100 | 86 | 86 |
| **13** | None | $-OCH_3$ | $Pb_4S_3Br_2$ | 100 | 93 | 93 |
| **14** | None | -Br | $CsPbBr_3$ $/Pb_4S_3Br_2$ | 92 | 61 | 66 |
| **15** | None | -Br | $CsPbBr_3$ | 99 | 53 | 53 |
| **16** | None | -Br | $Pb_4S_3Br_2$ | 93 | 44 | 47 |

* Obtained from three independent measurements.

To investigate the electronic properties' influence of *p*-substituents on thiophenol reactivity, we selected two representative substrates: 4-bromothiophenol and 4-methoxythiophenol, bearing either an electron-withdrawing (-Br) or electron-donor ($-OCH_3$) groups. As shown in Table 2, HSs gave again higher product yields than the single component NCs. Notably, with 4-methoxythiophenol the yield of **2b** was ca. 94 % (Figure S20, Table 2, entry 11). This result is consistent with the electron-donating nature of the $-OCH_3$ group, which increases the electron density of the aromatic ring, thereby enhancing its reactivity. Conversely, the -Br substituent, being more electron-withdrawing, makes the aromatic ring more electron-deficient, resulting in lower reactivity (Figure S21, Table 2, entry 14). Electron-donating groups stabilize the thiyl radical by reducing the spin density at the sulfur atom, explaining the highest yield with *p*-$OCH_3$-thiophenol.[26] Figure 3a shows a clear correlation between the substituent's sigma value ($\sigma_p$)[27] and product yield for $CsPbBr_3/Pb_4S_3Br_2$ HSs, consistent with electronic effects.

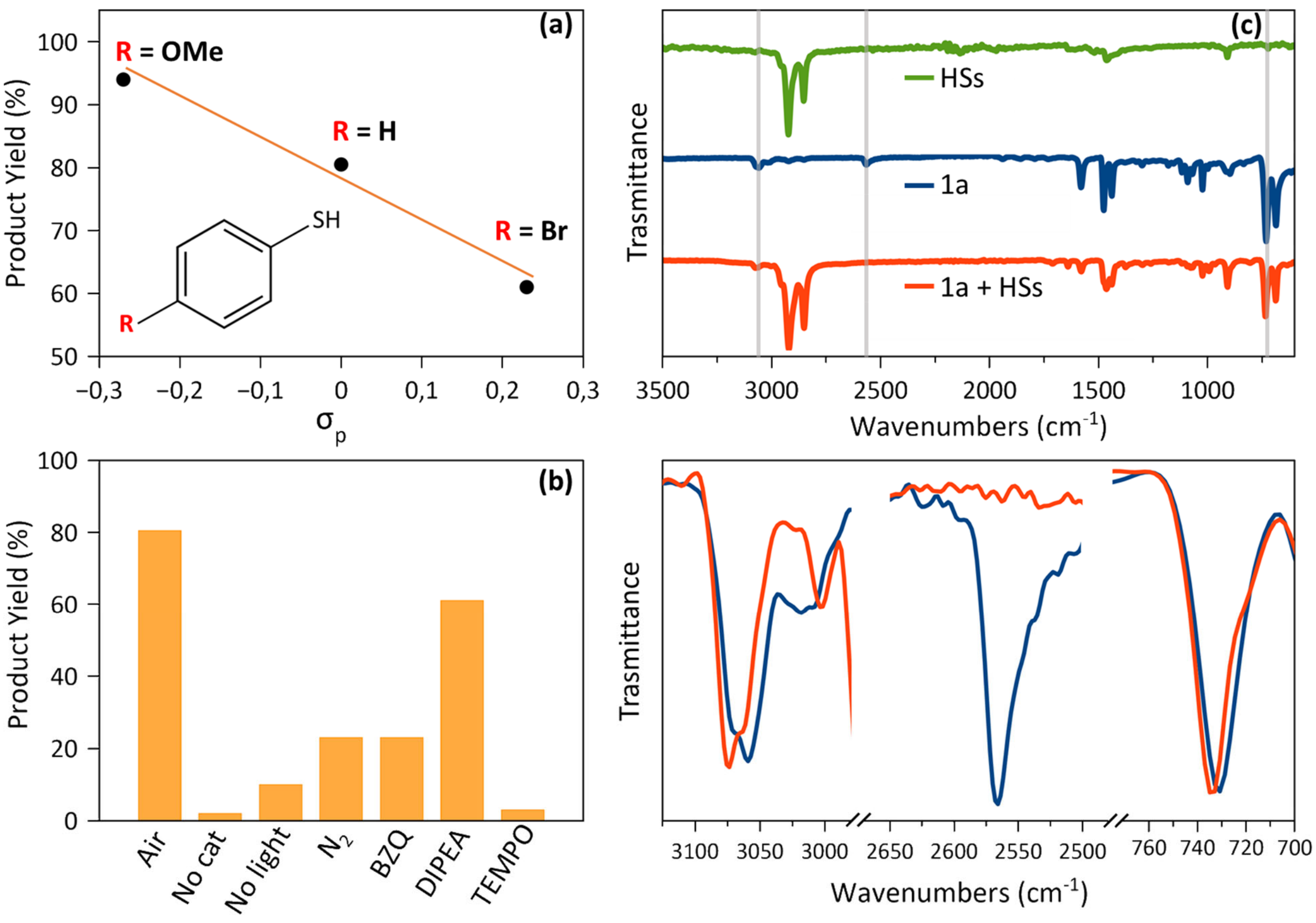

**Figure 3. Evaluation of different *p*-substituted thiophenols, different reaction conditions tested and interaction photocatalyst/substrate.** a) Correlation between the electronegativity of the R substituent on thiophenol and the disulfide yield. b) Product yield obtained from photocatalytic reactions tested under different conditions and in the presence of radical anion, holes, and radical scavengers. c) Top: FTIR spectra of $CsPbBr_3/Pb_4S_3Br_2$ HSs (green trace), 1a-thiophenol (blue trace), and the mixture between $CsPbBr_3/Pb_4S_3Br_2$ HSs with 1a-thiophenol (red trace). Bottom: magnification of selected FTIR stretching regions highlighting differences between thiophenol (blue trace) and the HSs–thiophenol mixture (red trace).

To elucidate the photocatalytic mechanism, several experiments with active trapping species were performed, employing 1,4 benzoquinone (BZQ)[28] , *N,N*-diisopropylethylamine (DIPEA),[29] and 2,2′,6,6′-tetramethylpiperidine-1-oxyl (TEMPO), as shown in Figure 3b and summarized in

Table S3. In the presence of BZQ, a well-known scavenger of superoxide anion ($O_2^{\cdot-}$), **1b** was obtained in 23% yield, which closely matches the yield observed under anaerobic conditions. Conversely, DIPEA, which is a hole ($h^+$) scavenger, led to a moderate reduction in yield (61%), suggesting that photogenerated holes also participate, albeit to a lesser extent, in the mechanism (as discussed below). When the reaction was performed in the presence of TEMPO as radical scavenger, under both aerobic and anaerobic conditions, only negligible amounts of product were obtained (Figure 3b), suggesting the formation of thiyl radicals (PhS). The detection of TEMPO-PhS adduct by GC-MS analyses (Figure S22) confirmed the production of thiyl radicals as key intermediates in disulfide formation.

The surface chemistry of the photocatalyst also plays an important role in the overall photocatalytic activity. The possible thiophenol adsorption onto the NCs' surface was investigated by Fourier-transform infrared spectroscopy (FTIR): the $CsPbBr_3/Pb_4S_3Br_2$-PhSH mixture spectrum was compared with those of the heterostructures and thiophenol alone (Figure 3c top). In the $CsPbBr_3/Pb_4S_3Br_2$–PhSH mixture the S–H band at 2565 $cm^{-1}$ is absent, indicating the interaction between thiophenol and the HSs. The C–S stretching shifts from 730 to 735 $cm^{-1}$, and the C–H stretching at 3050 $cm^{-1}$ shifts to higher energy ($\Delta\nu$ =14 $cm^{-1}$), reflecting changes in the chemical environment (Figure 3c down). Similar shifts were observed for mixtures of PhSH and "free standing" NCs and also in the case of *p*-$OCH_3$- and *p*-Br- thiophenol (Figures S23-26), in accordance with previous reports on organic molecules–perovskite interactions.[30] Collectively, these data support the adsorption of the different substrates onto the surface of the photocatalyst, which likely facilitates the photogenerated charge transfer from the HS to the substrate.

The proposed mechanism for the photooxidation of *p*-substituted thiophenols under aerobic conditions is represented in Figure 4. Once the contact between the photocatalyst and the substrate is established, upon visible-light illumination, the photogenerated holes migrate toward the $Pb_4S_3Br_2$ domain, while electrons are delocalized across the conduction band of the HS. This partial charge separation can reduce charge recombination and thus enhance photocatalytic activity compared to the individual components. Under aerobic conditions, two parallel pathways might produce thiyl radicals: i) photogenerated electrons from the conduction band of the HS reduce $O_2$ to form the superoxide anion ($O_2{\cdot}^-$), which reacts with thiophenol producing a thiyl radical (PhS·) and $HO_2^-$; and ii) photogenerated holes react with thiophenol, forming also a thiyl radical (PhS·) and a proton ($H^+$). Subsequently, two thiyl radicals can couple to form the target disulfide product. As the intermediate $HO_2^-$ is unstable, it can react with $H^+$ to produce $H_2O_2$. These finding confirm the cooperative activity of both photogenerated charge carriers (electron and hole) under aerobic conditions to produce thiyl radicals, thus improving the photocatalytic activity in the heterostructure. Under nitrogen atmosphere, there is only one pathway to produce thiyl radicals: the photogenerated electrons and holes interact with the thiophenol leading to the thiyl and hydrogen radicals, which after coupling produce the disulfide compound and molecular hydrogen ($H_2$, identified by GC Figure S27), respectively (Figure S28). Under both aerobic and anaerobic conditions, the coupling of two thiyl radicals leads to formation of a disulfide product, which subsequently detaches from the photocatalyst surface due to a lower affinity and diffuses into the solvent.

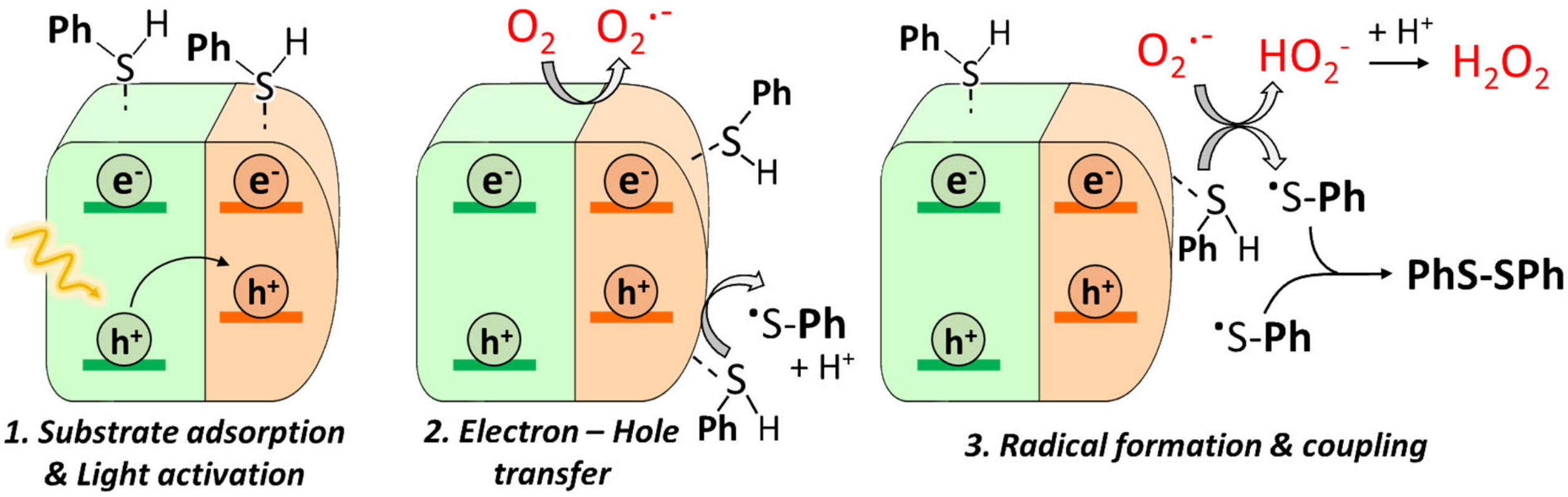

**Figure 4. Possible reaction mechanism.** Proposed photocatalytic mechanisms for the oxidative coupling of thiophenol under aerobic conditions.

The performance of the photocatalyst was evaluated after one photocatalytic cycle. XRD patterns recorded before and after the the reaction confirmed the preservation of crystallinity of the HSs (Figure S29). The absorption spectrum of $CsPbBr_3/Pb_4S_3Br_2$ HSs retained the perovskite band edge with a slight red shift and increased scattering (Figure S30), likely due to some nanocrystal aggregation caused by partial removal of surface ligands during the photoreaction, as previously observed in C–C coupling with $CsPbBr_3$ NCs.[25] TEM imaging of the crude solution (Figure S31) revealed significant aggregation in part of the sample, but the heterostructure composition was preserved.

To further validate our initial selection criteria for $CsPbBr_3/Pb_4S_3Br_2$ HSs, the photoreactions were performed also using Cl- (Table S4) and mixed I/Br-HSs (Table S5). Employing $CsPbCl_3/Pb_4S_3Cl_2$ HSs, the observed trend is consistent with the one found for the bromide analogue: the HSs outperform the individual NCs in both yield and selectivity. The slightly lower yield (70 % over 80%) obtained with the Cl-HSs can be attribuited to the type-I band alignment, which promotes non-radiative recombination and limits charge separation. The $CsPbI_3/Pb_4S_3Br_2$ HS, obtained *via* anion exchange, delivered the coupling product in 80% yield for **2a**, similar to that obtained with the bromide-based system, consistent with a band alignment. Nonetheless, in all cases, a reduction in the selectivity was observed.

In conclusion, we developed an optimized direct synthesis for semiconductor/semiconductor HSs ($CsPbBr_3/Pb_4S_3Br_2$, $CsPbCl_3/Pb_4S_3Cl_2$) and $CsPbI_3/Pb_4S_3Br_2$ *via* post-synthetic anion exchange, achieving tens-of-milligrams yields, above typical literature reports, and enabling extensive use in organic transformation photocatalysis. The chemical tunability of perovskite allowed modulation of the band alignment, systematically studied through optical and spectroscopic methods. Using *p*-substituted thiophenol coupling as a model reaction, the HSs outperformed the "free-standing" nanocrystals. Under aerobic conditions, the photogenerated electrons and holes produced thiyl radicals, enhancing the photocatalytic efficiency, exploiting the charge carrier separation at the heterojunction. This work establishes a versatile strategy for constructing semiconductor-semiconductor heterostructures. It demonstrates that type II heterojunction engineering, surface chemistry, and the cooperative participation of photogenerated charge carriers to produce intermediate radicals, is an effective approach for enhancing photocatalytic performance of perovskite NCs.

ASSOCIATED CONTENT

**Supporting Information.** Experimental methods and supporting figures including characterization of the photocatalyst by TEM images, ABS/PL spectra, XRD patterns, UPS spectra, XPS spectra, Tauc plot and TGA; product identification by GC-MS, FTIR spectra of different substrates and photocatalysts; tables of different photocatalysts tested.

AUTHOR INFORMATION

**Corresponding Authors**

Julia Pérez-Prieto − Institute of Molecular Science, University of Valencia, 46980 Paterna, Valencia, Spain; orcid.org/0000-0002-5833-341X; Email: julia.perez@uv.es

Ilka Kriegel − Department of Applied Science and Technology, Politecnico di Torino, 10129 Turin, Italy; Email: ilka.kriegel@polito.it

Liberato Manna − Nanochemistry Department, Italian Institute of Technology, 16163 Genova, Italy; orcid.org/0000-0003-4386-7985; Email: liberato.manna@iit.it

Raquel E. Galian − Institute of Molecular Science, University of Valencia, 46980 Paterna, Valencia, Spain; orcid.org/0000-0001-8703-4403; Email: raquel.galian@uv.es

**Authors**

Anna Cabona − Nanochemistry Department, Italian Institute of Technology, 16163 Genova, Italy; Department of Applied Science and Technology, Politecnico di Torino, 10129 Turin, Italy; orcid.org/0000-0002-5612-602X

Stefano Toso − Nanochemistry Department, Italian Institute of Technology, 16163 Genova, Italy; Lund University, Division of Chemical Physics, 221 00 Lund, Sweden; orcid.org/0000-0002-1621-5888

Alejandro Cortés-Villena − Institute of Molecular Science, University of Valencia, Paterna 46980 Valencia, Spain

Ignacio Rosa-Pardo − Institute of Molecular Science, University of Valencia, Paterna 46980 Valencia, Spain

Mirko Prato − Materials Characterization Facility, Italian Institute of Technology, 16163 Genova, Italy; orcid.org/0000-0002-2188-8059

Michele Ferri − Nanochemistry Department, Italian Institute of Technology, 16163 Genova, Italy; orcid.org/0000-0002-3862-6709

**Notes**

The authors declare no competing financial interest.

ACKNOWLEDGEMENT

A.C. and I.K. acknowledge funding from European Research Council through the ERC Starting Grant Light-DYNAMO (grant agreement n. 850875). This work was supported by Generalitat Valenciana (IDIFEDER/2018/064, IDIFEDER/2021/064, CIPROM/2022/57). It formed part of the Advanced Materials program (MFA/2022/051) and was supported by MCIN with funding from European Union NextGenerationEU (PRTR-C17.I1) and by Generalitat Valenciana. S. T. acknowledges the European Union's Horizon Europe research and innovation programme under the Marie Skłodowska-Curie Funding Program (Project SUPER-QD, Grant

Agreement No. 101148934). L. M. acknowledges funding from the European Research Council through the ERC Advanced Grant NEHA (grant agreement n. 101095974).

# SUPPORTING INFORMATION

## EXPERIMENTAL METHODS

**Chemicals.** 1-Octadecene (ODE, tech, 90%), oleic acid (OA, tech, 90%), oleylamine (OLA, tech, 70%), lead(II) bromide ($PbBr_2$, 98%), lead(II) chloride ($PbCl_2$, 98%), lead(II) iodide ($PbI_2$, 98%), cesium carbonate ($Cs_2CO_3$, 99%), 1-dodecanethiol (DDT, 99.9%), sulfur powder (S, 99.99%), lead acetate trihydrate ($Pb(OAc)_2{\cdot}3H_2O$, 99.99%), dimethyl sulfoxide (DMSO, 99.5%), were purchased from Sigma-Aldrich. All reagents were used as received without any further experimental purification.

**Preparation of Cs-oleate Stock Solution.** In a typical synthesis, 120 mg (0,37 mmol) of $Cs_2CO_3$ were mixed with 1.75 mL of OA and 15 mL ODE in 50 mL three-neck flask, dried for 1h at 110°C and then heated under $N_2$ to 150°C until the solution turned clear. Thereafter, the solution was transferred into $N_2$-filled glass vials. The solution is solid at room temperature and needs to be pre-heated before using it.

**Preparation of $Pb(OA)_2$ Stock Solution.** $Pb(OAc)_2{\cdot}3H_2O$ powder (0.38 g, 1 mmol) and OA (950 uL) were mixed with ODE (9.35 mL) in a 50 mL three-neck round-bottom flask. The reaction mixture was degassed under vacuum for 1 h at 110°C and then heated under $N_2$ to 150°C until all $Pb(OAc)_2{\cdot}3H_2O$ reacted with OA. Thereafter, the solution was cooled to room temperature (25 °C) and transferred into $N_2$-filled glass vials.

**Preparation of S-ODE Stock Solution.** 1 mmol of S powder was mixed with 10 mL ODE (predegassed at 120°C for an hour) in a 20 mL glass. Then, the resulting mixture was sonicated until the complete dissolution of S (ca. 20 min).

**Synthesis of $CsPbBr_3/Pb_4S_3Br_2$ HSs.** A typical HS synthesis involves two steps, performed consecutively: synthesis of $CsPbBr_3$ nanocrystals and growth of $CsPbBr_3/Pb_4S_3Br_3$ HSs. In the first step, 72 mg of $PbBr_2$ are mixed with 5 ml of octadecene, 50 µL of oleic acid, and 500 µL of oleylamine inside a 20 mL glass vial, under nitrogen atmosphere. The mixture is heated to 170°C to achieve the full solubilization of the powder, followed by the injection at 150°C of 500 µL of a Cs-oleate solution is injected to initiate the growth of perovskite nanocrystals (NCs). After 5 seconds, the reaction is quenched by immersion in a water bath. The reaction batch is then stirred for 5 minutes at room temperature without performing any purification. Subsequently, the vial is reheated to 220°C, and the precursors needed for the growth of the chalcohalide are injected: 800 µL of lead oleate, 32 µL of 1-dodecanethiol and 480 µL of sulfur-ODE. The reaction is allowed to proceed for 30 seconds before being quenched in an ice water bath. The crude solution is then centrifuged at 6000 rpm for 5 minutes to isolate the NCs. The precipitate is collected in hexane, followed by a second centrifugation at a 2000 rpm for 2 minutes to remove larger perovskite NCs that may have formed during the re-heating step. The precipitate is discarded, and the supernatant is kept for further use.

**Synthesis of $CsPbCl_3/Pb_4S_3Cl_2$ HSs.** The synthesis of Cl-based HSs follows the same protocol adopted for Br-based HSs, with the following adaptations: 1) 53 mg of $PbCl_2$ were used instead of $PbBr_2$ and the amount of oleic acid was increased to 500 µL to facilitate the dissolution of $PbCl_2$, 2) the first injection temperature was raised to 185°C, as lower temperature (150°C) resulted in a high number of platelets instead of nanoparticles. The procedure to obtain and purify the HSs is the same as the bromide-based sample.

**Isolation of $Pb_4S_3X_2$ nanocrystals.** Pure $Pb_4S_3X_2$ nanocrystals can be obtained by etching the perovskite domains of the corresponding $CsPbX_3/Pb_4S_3X_2$ HSs. First, a solution of HSs in hexane is centrifuged at 2000 rpm for 2 minutes to remove aggregates, which ensures uniform and effective etching. Subsequently, 1 mL of the HS solution in hexane is mixed with 1 mL of dimethyl sulfoxide and 60 µL of oleylamine. The mixture is vortexed for approximately 60 seconds. After a few minutes, phase separation occurs due to the mutual insolubility and density difference between hexane and dimethyl sulfoxide. The upper hexane layer, which appears red due to the presence of chalcohalide nanoparticles, is collected. After that, 4 mL of ethyl acetate and 60 µL of oleic acid are added to the hexane fraction to precipitate the particles, which are collected by centrifugation (6000 rpm for 5 minutes) and then redispersed in hexane or toluene for further use.

**Anion exchange.** The anion exchange reactions were performed under ambient conditions following previously reported methods.[1] First, a $PbI_2$ stock solution was prepared: 2 mmol of PbI2, 5 mL of OA and 5 mL of OLAm were mixed with 30 mL of ODE in a 100 mL three-neck flask. The reaction mixture was dried/degassed under vacuum for 30 min at 110°C. Then, the flask was filled with $N_2$, and the temperature was raised to 150°C. After complete dissolution of $PbI_2$ salt, the solution turned yellow and then cooled down to room temperature (25°C) and transferred into a $N_2$-filled glass vial. Then, 4 mL of the $CsPbBr_3$–$Pb_4S_3Br_2$ HSs in toluene was added into a 10 mL glass vial, and different amounts of $PbI_2$ stock solution (ranging from 300 uL to 900 µL) were added under vigorous stirring at RT for at least 30 min. Thereafter, the HSs were collected by centrifugation at 6000 rpm for 5 min and redispersed in toluene for further use.

**Optical Properties.** Absorbance spectra were measured with a Cary300 UV–Vis absorption spectrophotometer, while PL spectra were recorded by a Varian Cary Eclipse spectrophotometer using an excitation wavelength at 350 nm for the bromide-based samples and 300 for the chloride-based ones. The NC solutions were diluted in hexane in quartz cuvettes (path length = 1 cm) to a maximum optical density below 1.0.

**Transmission Electron Microscopy.** Bright Field Transmission Electron Microscopy (BF-TEM) measurements of the NCs were performed using a JEOL JEM-1011 with a W thermionic source at an acceleration voltage of 100 kV. The highly diluted NC solution was drop-cast onto copper grids (200 mesh) with carbon film, and the solvent was then allowed to evaporate in a vapor-controlled environment. The longitudinal and lateral dimensions were assessed through statistical analysis of TEM images of several hundred NCs using the ImageJ software.

**X-ray Powder Diffraction.** Characterization by X-ray Powder Diffraction (XRD) was performed by employing a PANalytical Empyrean X-ray diffractometer using a 1.8 kV Cu Kα ceramic X-ray tube operating at 45 kV and 40 mA and detected by a PIXcel3D 2 × 2 area detector. Samples were prepared by drop-casting highly concentrated solutions on zero-diffraction silicon substrates. All diffraction patterns were acquired at room temperature under ambient conditions. Data analysis was performed using the HighScore 4.9 software from PANalytical.

**X-ray Photoelectron Spectroscopy.** XPS specimens were prepared by drop casting a few microliters of the sample dispersions onto freshly cleaved highly oriented pyrolytic graphite (HOPG, ZYB grade) substrates. XPS data were acquired using a Kratos Axis UltraDLD spectrometer, equipped with a monochromatic Al Kα source operated at 20 mA and 15 kV. Wide scans were acquired at a pass energy of 160 eV, energy step of 1 eV, over an analysis area of 300 x 700 μm2. Spectra have been charged corrected, the lowest binding-energy component of the carbon 1s spectrum was set at 284.8 eV. The collected data were then analyzed with CasaXPS software (version 2.3.24).[2]

**Ultraviolet Photoelectron Spectroscopy.** UPS was carried out on the same spectrometer, using a He I (21.22 eV) discharge lamp, on an area of 55 μm in diameter, at a pass energy of 10 eV and with a dwell time of 100 ms. The work function (that is, the position of the Fermi level with respect to the vacuum level) was measured from the threshold energy for the emission of secondary electrons during He I excitation. A −9.0 V bias was applied to the sample to precisely determine the low-kinetic-energy cutoff, as discussed in ref[3]. The position of the cutoff was then estimated with CasaXPS software, using the “Edge Up” background function for the energy interval around the cutoff. Then, the position of the VBM versus the vacuum level was estimated by measuring its distance from the Fermi level, focusing on the high-kinetic energy (i.e., low-binding energy) cutoff region and using the “Edge Down” background function in CasaXPS software.

**Ambient Pressure Photoemission Spectroscopy.** The ionization (I) energy or valence band (VB) energy of the studied materials was determined using an Ambient Pressure Photoemission Spectroscopy (APS) system (model APS02, KP Technology Ltd, Highlands and Islands, United Kingdom). The APS02 system was operated with an incident photon energy range from 4.56 to 6.89 eV, generated using a tunable UV light source of Deuterium. The samples for APS measurements were prepared by drop-casting thin films of the materials onto ITO-coated glass substrates, which provide a conductive and transparent support compatible with photoemission studies. The APS chamber was continuously purged with nitrogen gas during operation to effectively suppress ozone formation throughout the measurement. The onset of the photoemission signal with the background was used to determine the VB energy.

**Electrochemistry.** Electrochemical characterization was performed on an Autolab 128N potentiostat/galvanostat using a three-electrode system in a glass beaker. Cyclic voltammetry (CV) experiments were carried out in 0.1 M tetrabutylammonium tetrafluoroborate ($TBABF_4$) solution in a mixture of anhydrous ACN:toluene (1:3 v/v) with a thiophenol concentration of 0.12 M. A glassy carbon, Pt wire and Ag/AgCl were used as working, counter and reference electrodes, respectively. The measurements were performed at room temperature (298 ± 1 K) and under aerated conditions, with a scan rate of 100 mV/s. To determine the energy of the highest-occupied molecular orbital (HOMO) level, the anodic peak potential (in volts) was referred to the Fc/Fc+ redox couple using 0.5 mM solutions of ferrocene in 0.1 M $TBABF_4$ in the same electrolyte.

**Thermogravimetric analysis.** Thermogravimetric analysis (TGA) of the NCs was carried out with a Mettler Toledo TGA/SDTA851e/SF/1100 apparatus in the 25-800°C temperature range under a 10°C min' scan rate under nitrogen atmosphere.

**Attenuated total reflectance-Fourier transform infrared spectroscopy.** To ascertain the substrate approach to the photocatalysts NCs surface, the NCs and the substrates were mixed at the same concentration used in the photocatalytic reactions and the mixture was stirred to ensure the interaction between them. Then, the ATR-FTIR analyses were performed in a Bruker alpha II spectrometer by drop-casting the colloidal mixtures.

**Photocatalytic Experiments.** For the photocatalytic experiments, 1 mg of the photocatalyst was mixed with the stock solution of the substrate (thiophenol/ 1,4-bromothiophenol/ 1,4-methoxythiophenol) in 1 mL of solvent (cyclohexane) to reach the final concentration of 34 mM. Subsequently, the photoreactions were performed in a 10 mL gastight crimped vials under stirring (160 rpm) using an orbital shaker in a photoreactor with blue LEDs (450/405 nm) for 90 min at 20°C. After the reaction time, the crude solution was centrifuged at 12500 rpm for 15 minutes, the precipitate was collected in 1 mL of hexane for further use while the supernatant was analyzed by GC-MS to determine the conversion of the substrate and the yield of product. For the GC analysis the biphenyl (internal standard) stock solution was added to the supernatant to have a final concentration of 150 mM. Biphenyl was solubilized in ethyl acetate.

*Control experiments of the photocatalytic coupling of thiophenol.* All the reactions have been performed in the same conditions reported for the photocatalytic reactions but in the presence of the scavengers. Regarding the control experiments, 10 equivalents of 1,4-benzoquinone, *N,N*-Diisopropylethylamine (DIPEA) and 2,2′,6,6′-tetramethylpiperidine-1-oxyl (TEMPO) were added as scavengers.

*Control experiments in nitrogen atmosphere for the detection of $H_2$.* All the reactions under a nitrogen atmosphere for $H_2$ detection were performed using strictly anhydrous solvents inside a glovebox to prevent any moisture contamination from the atmosphere.

*Photocatalytic Setup.* Light source: the reactions were performed using (1= 450‡10 m or 1= 405‡10 m) LUXEON LED, mounted on a 10mm Square Saber - 1030 mW@700mA as a light source. Temperature Control: reaction temperature was controlled by a high-precision thermoregulation Hubber K6 cryostat. Likewise, to guarantee stable irradiation the temperature of the LEDs was set up at 20°C. The reactions have been carried out in an in-house parallel High Throughput Screening (HTS) (see Picture 1) photoreactor with capacity to set up to 25 reactions with different excitation wavelengths, respectively, under high-intensity irradiation. These unique HTS platforms allow for tight control of the light intensity and the temperature of the reactions. The 25-positions photoreactor is operable at 1-15 mL reaction volumes for each reaction.

**Gas chromatography – mass spectroscopy.** The work-up of the reaction was performed by adding biphenyl (IS) to the crude. Then, 0.1 mL of the mixture was diluted with 0.9 mL of cyclohexane. The solution was injected into the GC-MS (Agilent 7890B - 5977A) and the products were identified according to the retention time and mass. A calibration curve was established using different concentrations of analytes ($C_{AN}$) relative to a fixed concentration of the internal standard ($C_{IS}$ = 150 mM). The GC signals were then analyzed by comparing the peak areas of the analytes with that of the internal standard. The ratio between the product area and the internal standard allowed us to quantify the product concentrations in the reactions. This calibration plot was subsequently used to quantify the obtained products.

The following equations were used to calculate the product yield (1), substrate conversion (2) and the selectivity (3):

$\boldsymbol{Product\ Yield}$ (%) = $[\boldsymbol{Pro(AN)}]/[\boldsymbol{Initial\ substrate}]\ \boldsymbol{x100}$% (1)

$\boldsymbol{Conversion}$ (%) = $\boldsymbol{100} - [\boldsymbol{Final\ substrate}]/[\boldsymbol{Initial\ substrate}]\ \boldsymbol{x100}$% (2)

$\boldsymbol{Selectivity}$ (%) = $[\boldsymbol{Product\ Yield}]/[\boldsymbol{Conversion}]\boldsymbol{x100}$% (3)

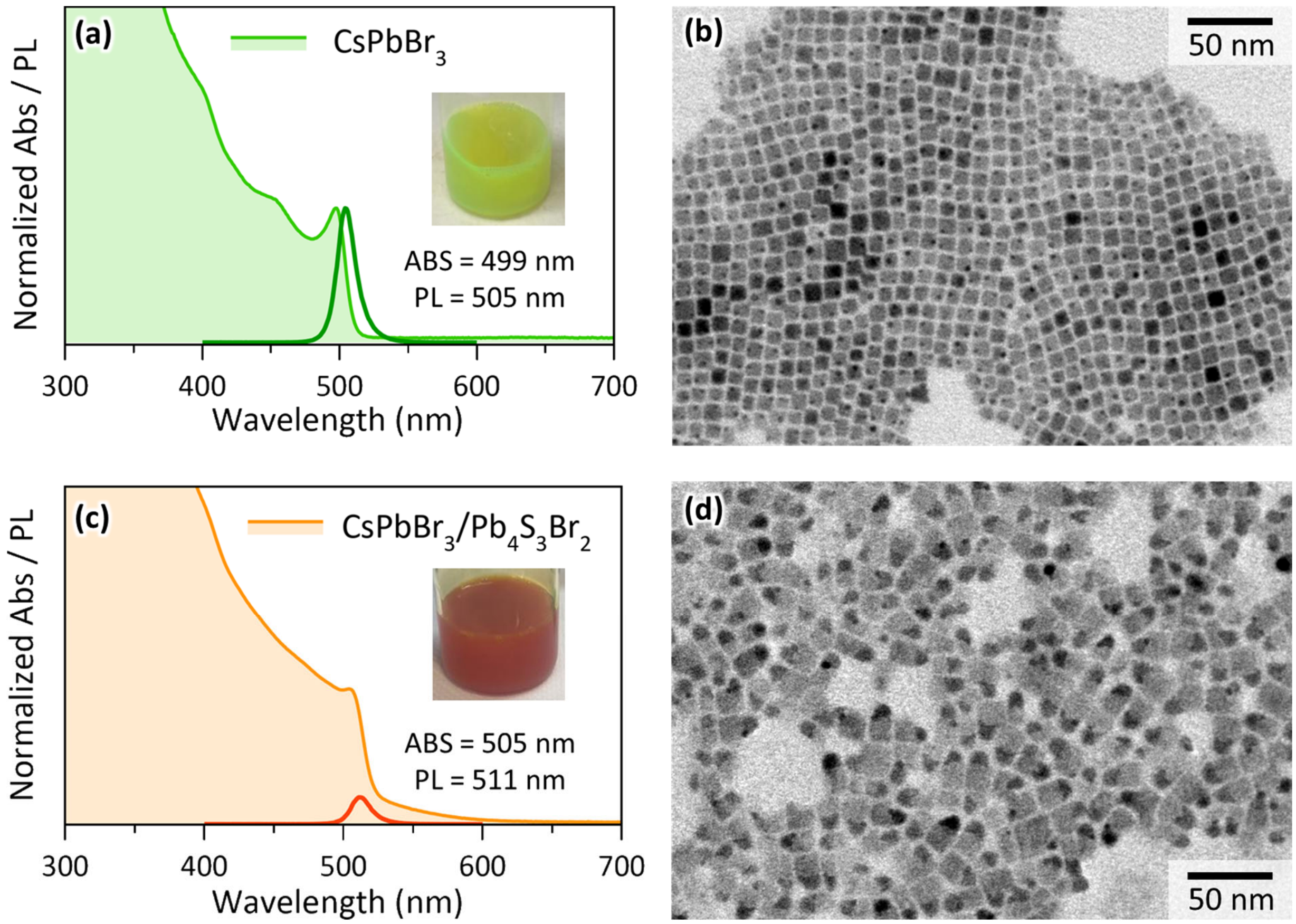

**Figure S1. Optical characterization and TEM images of $CsPbBr_3$ and $CsPbBr_3/Pb_4S_3Br_2$ heterostructures.** a) Absorption and PL spectra of $CsPbBr_3$ NCs synthetized at 150 °C prior to the growth of the chalcohalide domains. b) TEM images of the same $CsPbBr_3$ nanocrystals. c) Absorption and PL spectra of $CsPbBr_3/Pb_4S_3Br_2$ heterostructures after the growth of the chalcohalide domains. The PL was acquired with the same dilution factor and conditions of panel (a) and shows a marked reduction in intensity compatible with the formation of heterostructures, which are known to suppress the photoluminescence of the $CsPbBr_3$ domain. d) TEM images of $CsPbBr_3/Pb_4S_3Br_2$ heterostructures. Comparing the position of the absorption features, and TEM images before and after the growth of the chalcohalide, highlights some enlargement of the $CsPbBr_3$ domains, whose average size increases from 9.1 nm to 11.7 nm. This is likely due to the reaction of some of the additional Pb-oleate with unreacted $Cs^+$ in present in the reaction environment.

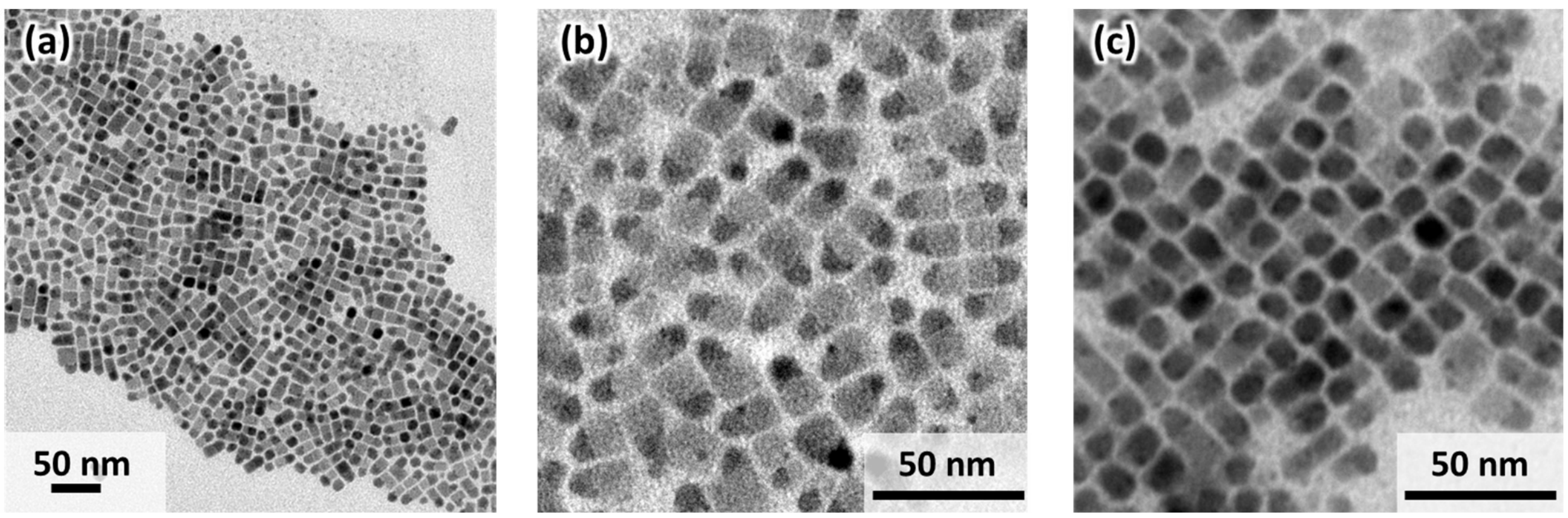

**Figure S2. TEM images of $CsPbBr_3/Pb_4S_3Br_2$ heterostructures at different magnifications.** The different projections visible in panel (a) allow to see some heterostructures lying flat on the TEM grid (b) (rectangular cross-section) as well as some standing upright (c) (squared cross-section).

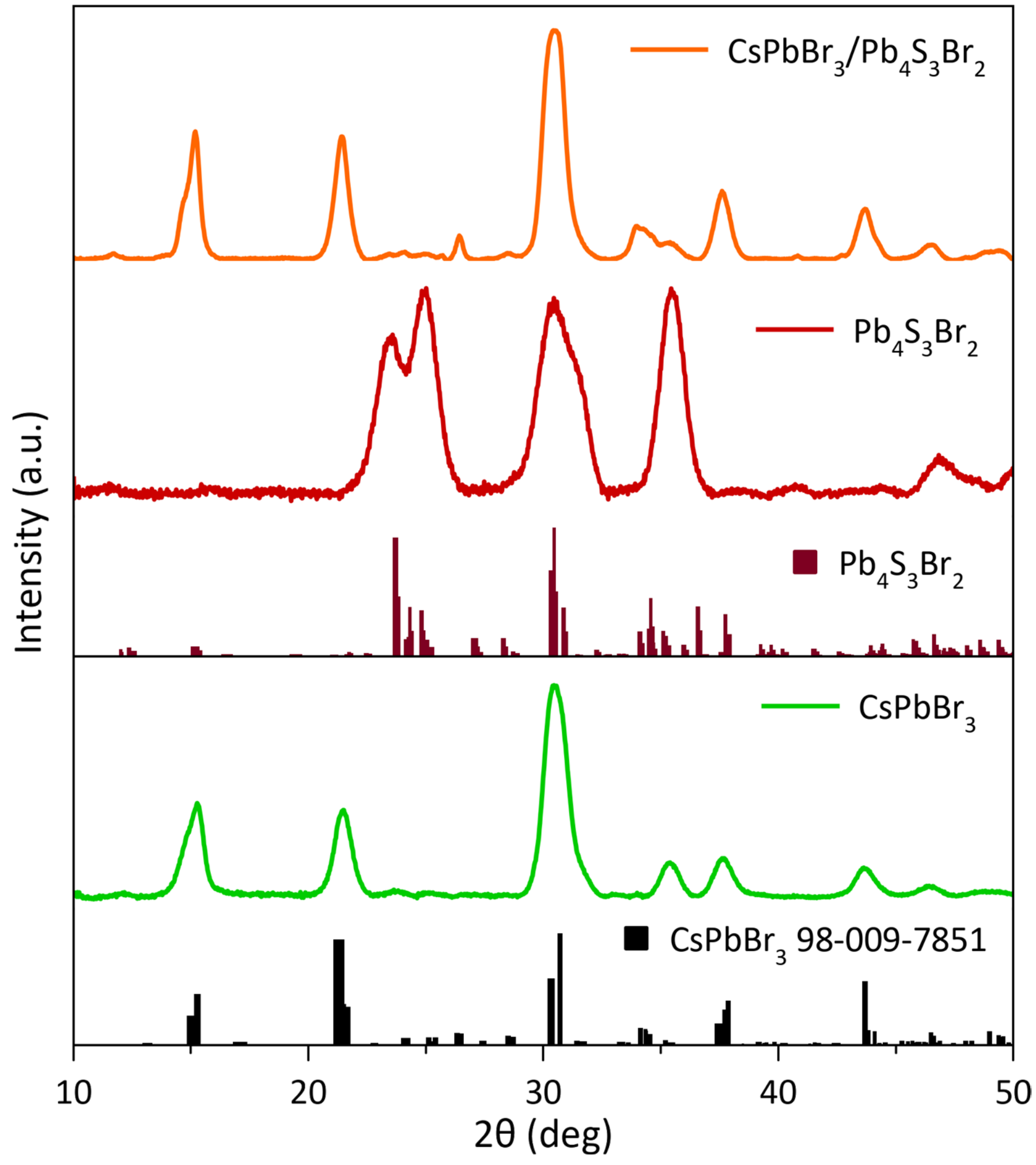

**Figure S3. XRD patterns $CsPbBr_3$ and $Pb_4S_3Br_2$ free-standing nanocrystals and heterostructures.** XRD pattern of $CsPbBr_3/Pb_4S_3Br_2$ HSs (orange line), $Pb_4S_3Br_2$ NCs (red line)

and reference pattern (bordeaux line), $CsPbBr_3$ (green line) and $CsPbBr_3$ reference (black line). The contribution of $Pb_4S_3Br_2$ to the heterostructures pattern is minor due to the much lower volume fraction compared to $CsPbBr_3$ and to the spread of diffracted intensities over more peaks due to the lower symmetry (although both materials crystallize in the Pnma space group, the pseudocubic symmetry of $CsPbBr_3$ causes many broad reflections to stack onto each other).

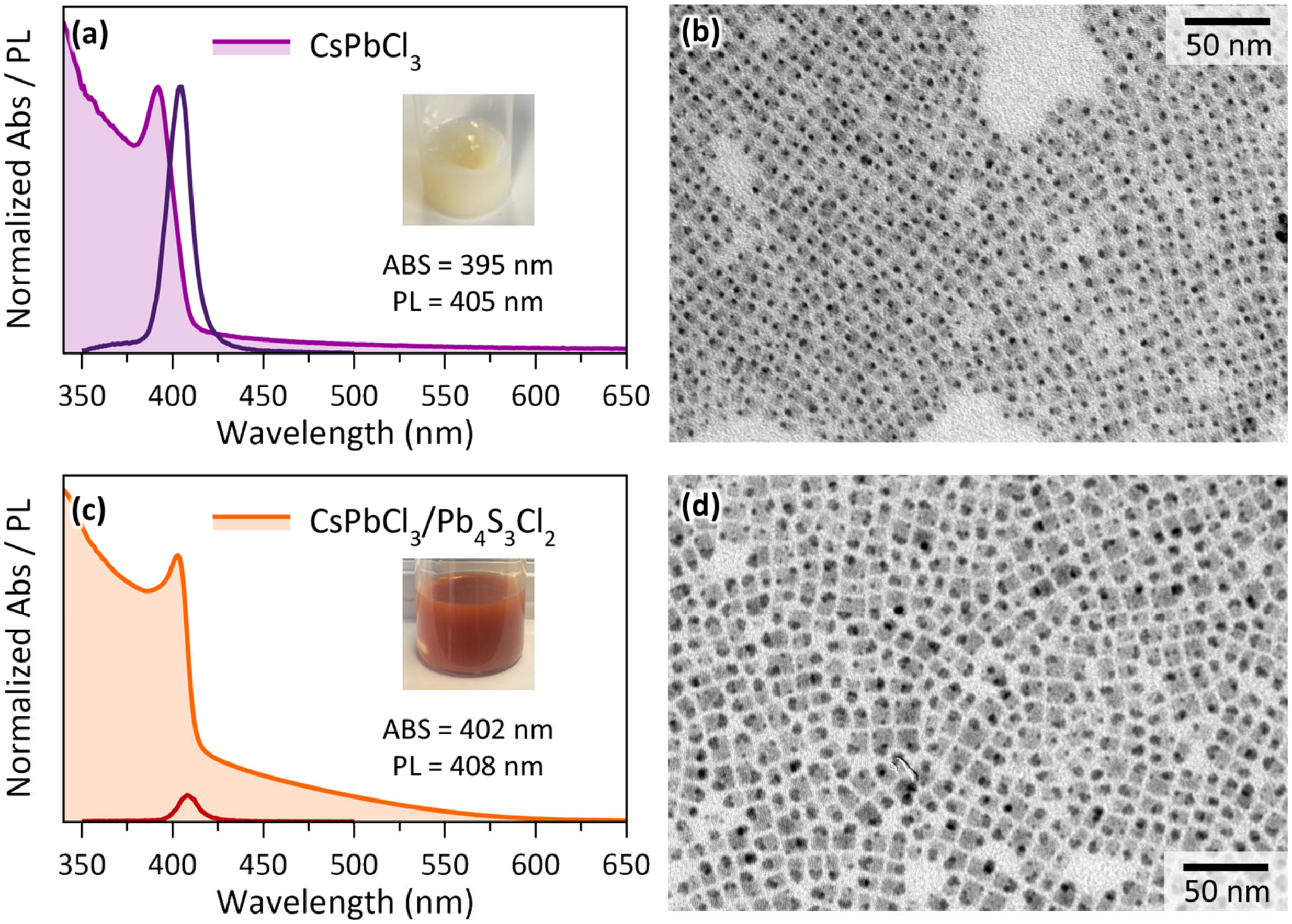

**Figure S4. Optical characterization and TEM images of $CsPbCl_3$ nanocrystals and $CsPbCl_3/Pb_4S_3Cl_2$ heterostructures.** a) Absorption and photoluminescence spectra of $CsPbCl_3$ nanocrystals synthetized at 185 °C. b) TEM images of the same $CsPbCl_3$ nanocrystals. c) Absorption and photoluminescence spectra of $CsPbCl_3/Pb_4S_3Cl_2$ heterostructures. d) TEM images of the same $CsPbBr_3/Pb_4S_3Cl_2$ heterostructures.

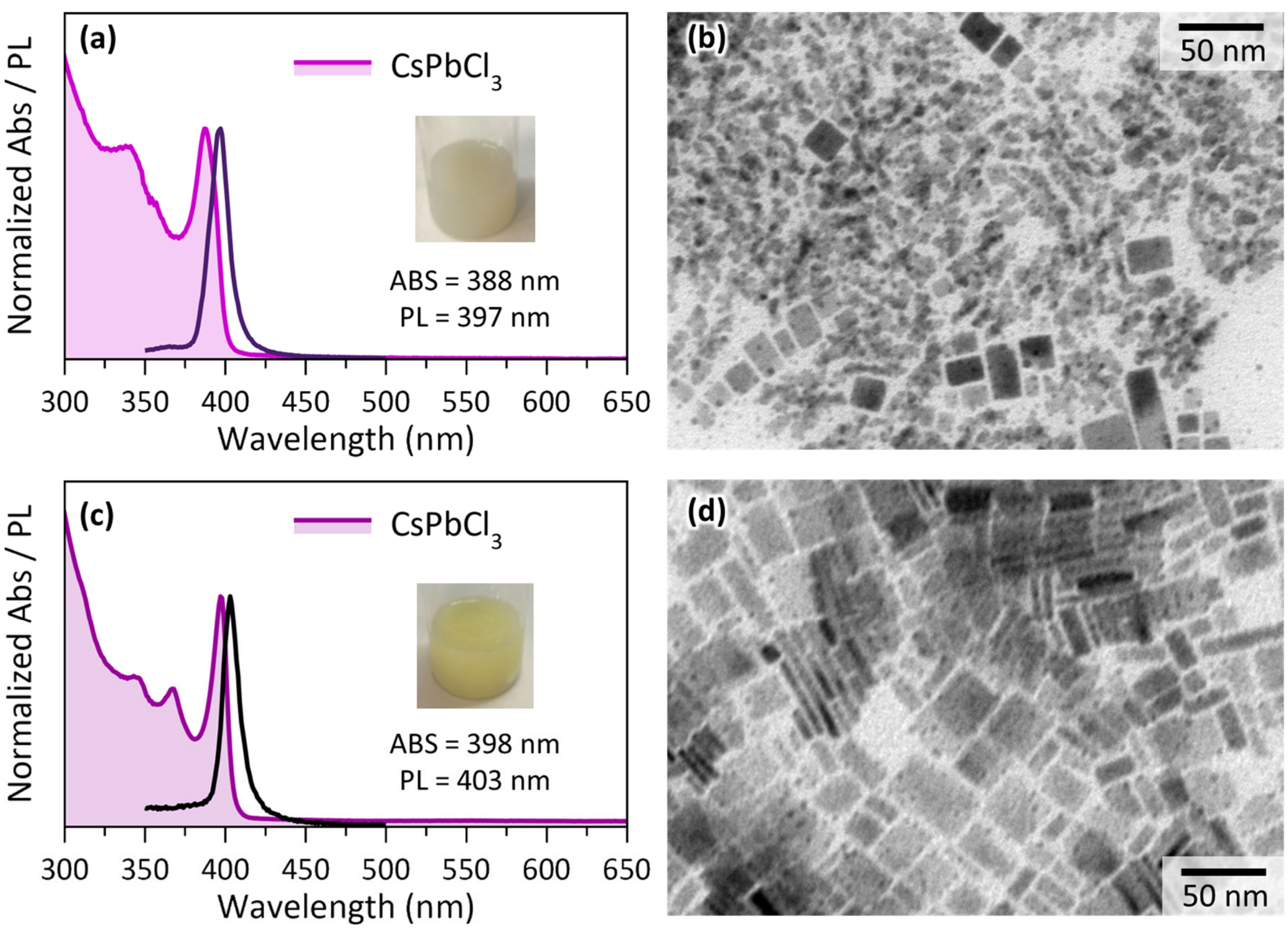

**Figure S5. Optimization of the $CsPbCl_3$ nanocrystal seeds for heterostructure growth.** a) Absorption and photoluminescence spectra of $CsPbCl_3$ nanocrystals synthetized at 150 °C and with the use of TOP (trioctylphosphine). b) TEM image of the same $CsPbCl_3$ nanocrystals. c) Absorption and photoluminescence spectra of heterostructures growth tests performed on the $CsPbCl_3$ nanocrystals (panel b). d) TEM image of the same test, no heterostructures are found under these conditions.

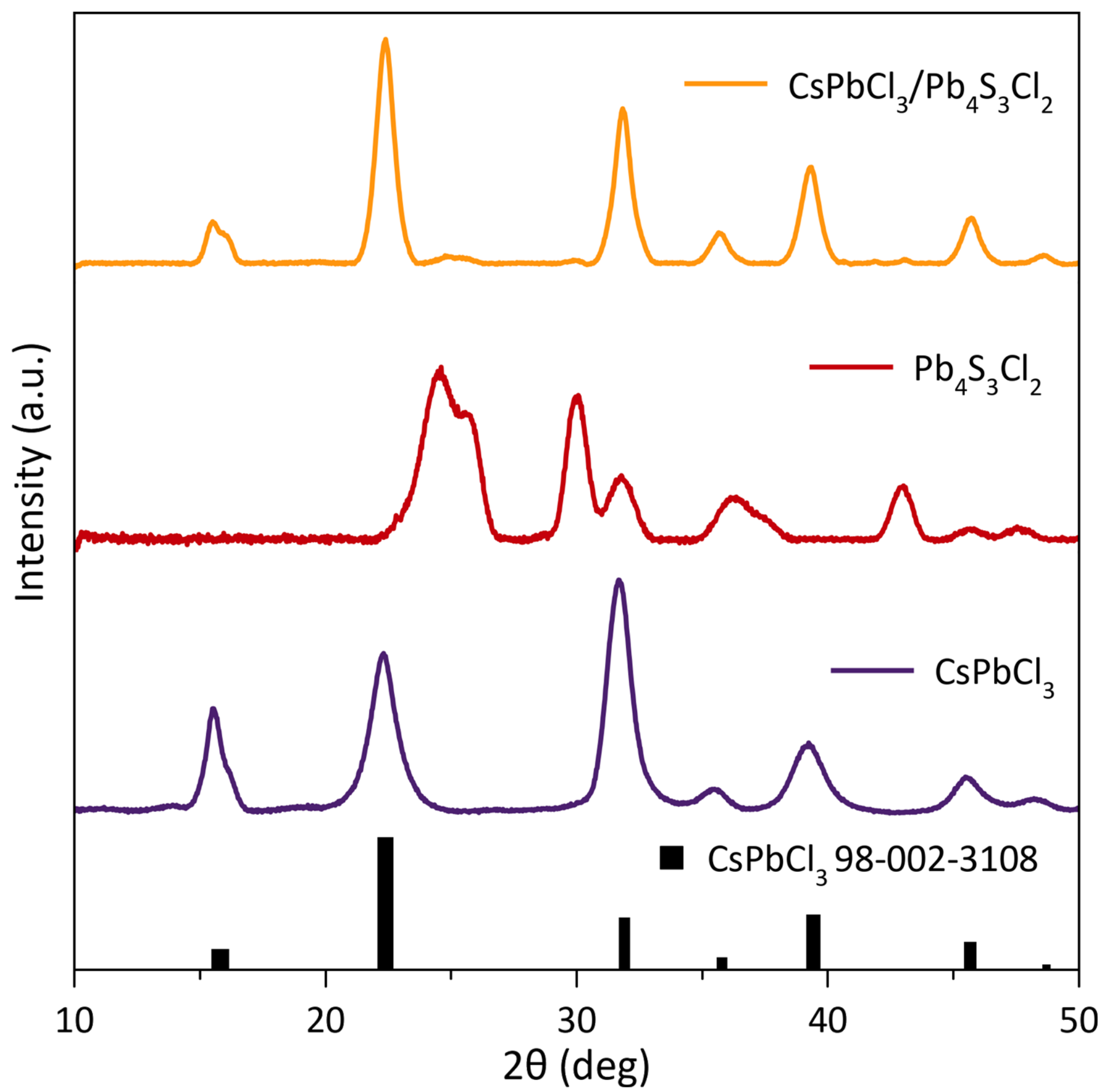

**Figure S6. XRD patterns $CsPbCl_3$ and $Pb_4S_3Cl_2$ free-standing nanocrystals and heterostructures.** XRD pattern of $CsPbCl_3$/$Pb_4S_3Cl_2$ HSs (orange line), $Pb_4S_3Cl_2$ NCs (red line) and $CsPbCl_3$ (purple line) $CsPbCl_3$ reference (black line). The contribution of $Pb_4S_3Cl_2$ to the heterostructures pattern is minor due to the much lower volume fraction compared to $CsPbCl_3$ and to the spread of diffracted intensities over more peaks due to the lower symmetry.

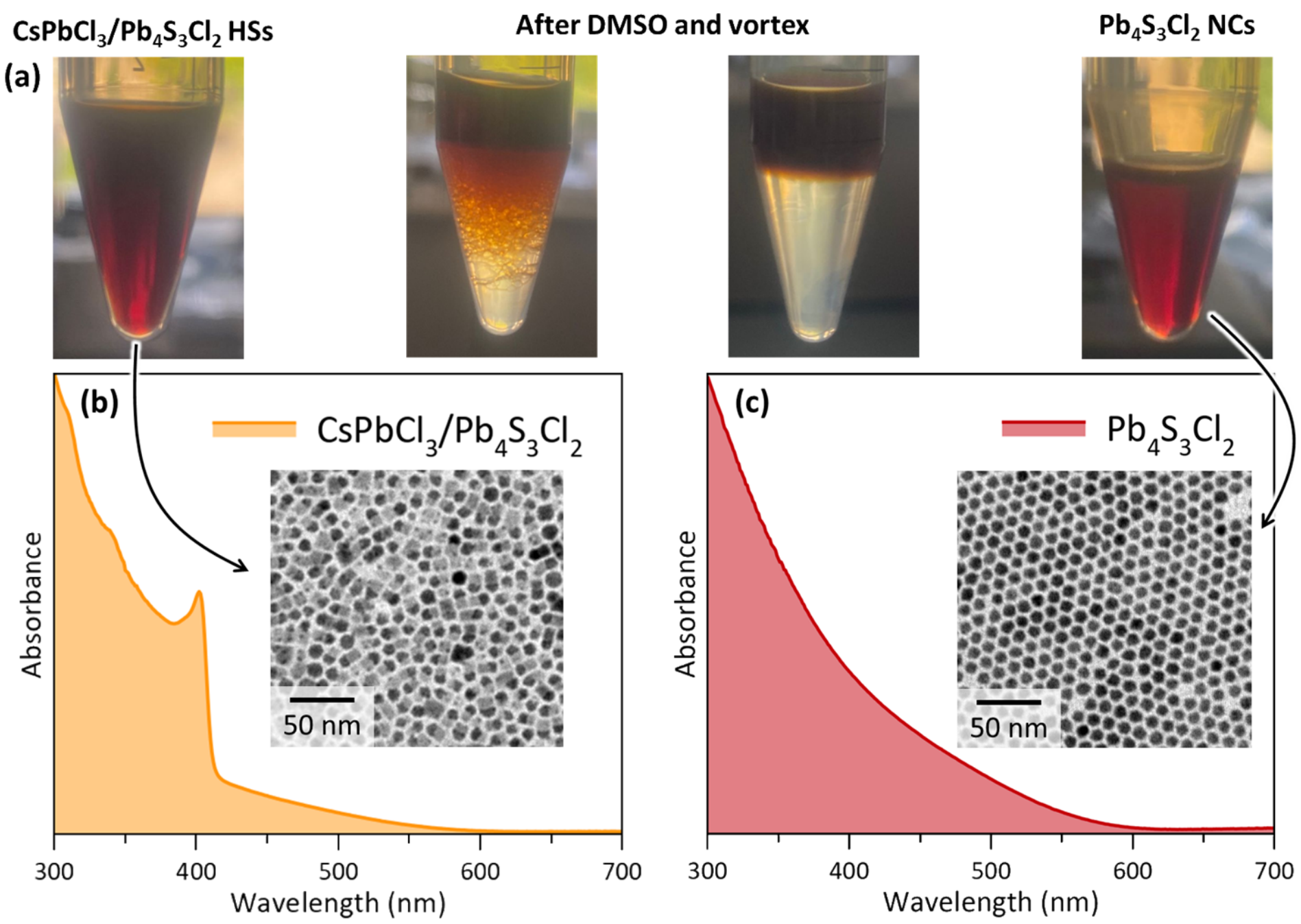

**Figure S7. Photographs of $CsPbCl_3/Pb_4S_3Cl_2$ HSs samples before and after selective perovskites etching, and corresponding optical absorbance and TEM images.** a) Photographs of heterostructures sample before and after the addition of DMSO to etch the perovskites domain. b) Absorbance spectrum of $CsPbCl_3/Pb_4S_3Cl_2$ HSs before addition of DMSO, inset: TEM images of HSs. c) Absorbance of the resulting $Pb_4S_3Cl_2$ NCs, inset: TEM images of the NCs.

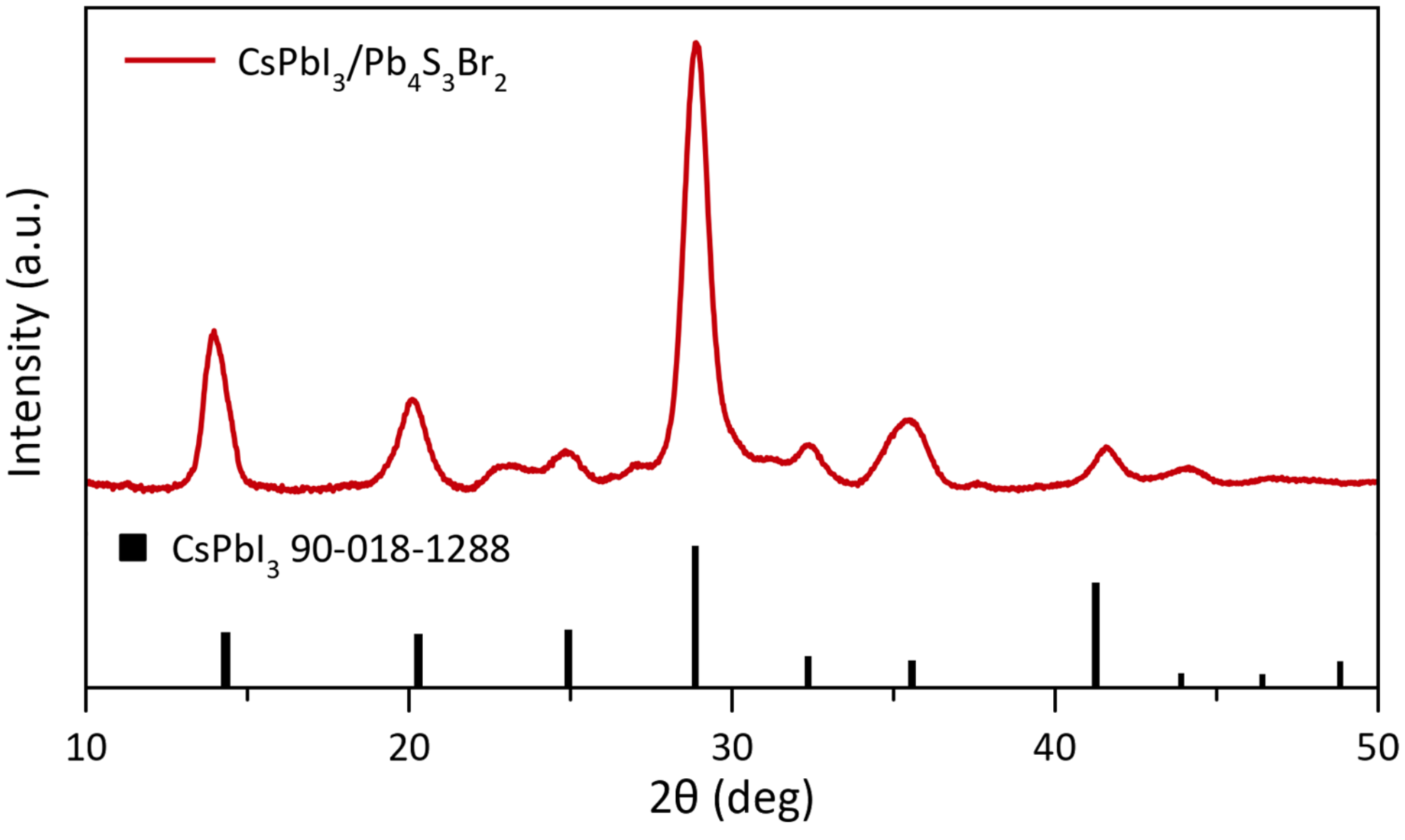

**Figure S8. XRD patterns of $CsPbI_3/Pb_4S_3Br_2$ HS.** XRD pattern of $CsPbI_3/Pb_4S_3Br_2$ HSs (red) after anion exchange, $CsPbI_3$ reference (black line). The contribution of $Pb_4S_3Br_2$ to the heterostructures pattern is minor due to the much lower volume fraction compared to $CsPbI_3$ and to the spread of diffracted intensities over more peaks due to the lower symmetry.

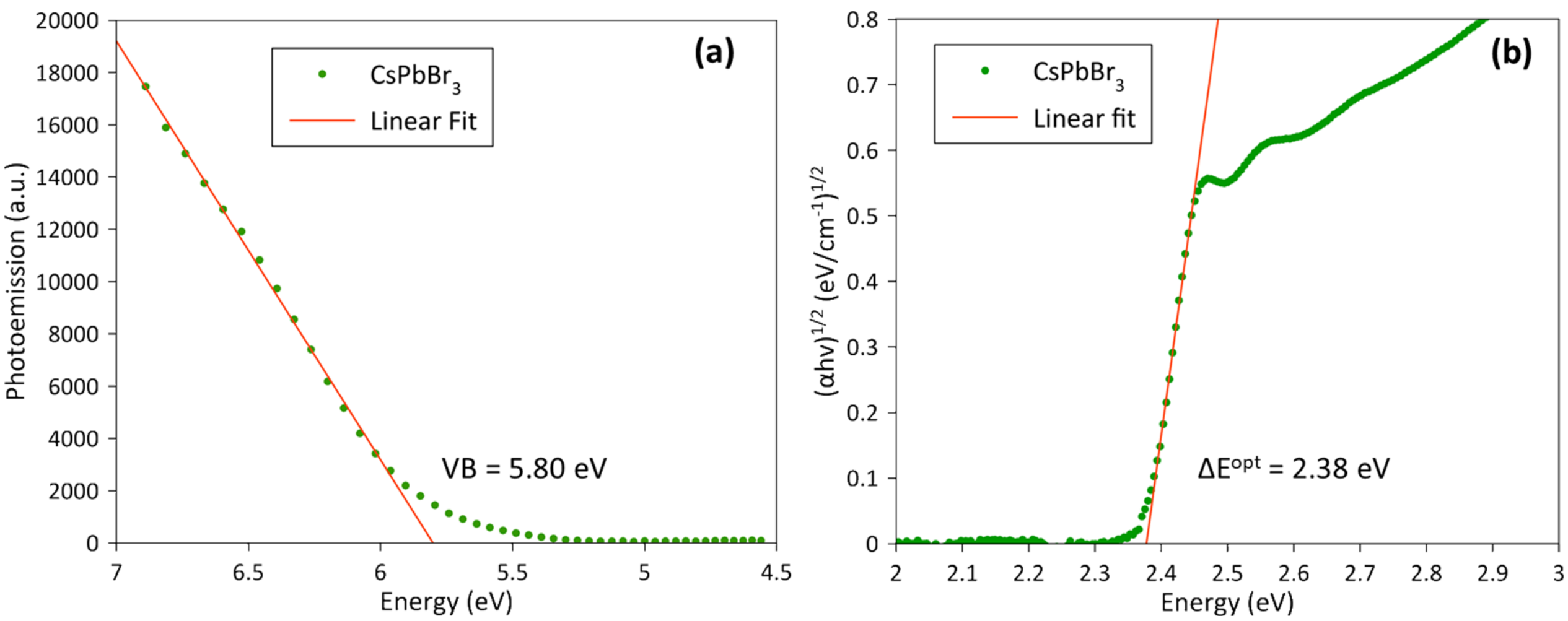

**Figure S9. Ambient Pressure Photoemission Spectroscopy and Tauc Plot of $CsPbBr_3$.** a) Photoemission of $CsPbBr_3$, used to determine the valence band of the material. b) Tauc plot from the absorbance spectrum of $CsPbBr_3$, used to determine the optical band gap of the material.

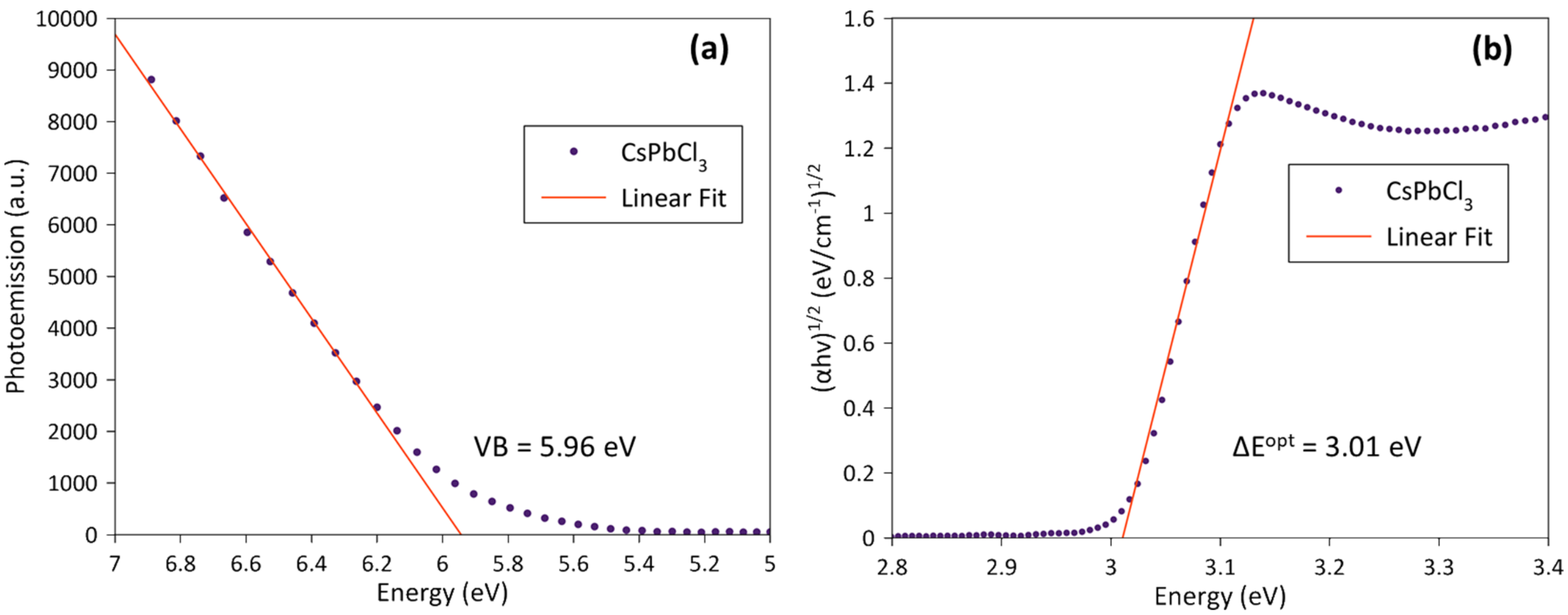

**Figure S10. Ambient Pressure Photoemission Spectroscopy and Tauc Plot of $CsPbCl_3$.** a) Photoemission of $CsPbCl_3$, used to determine the valence bad of the material. b) Tauc plot from the absorbance spectrum of $CsPbCl_3$, used to determine the optical band gap of the material.

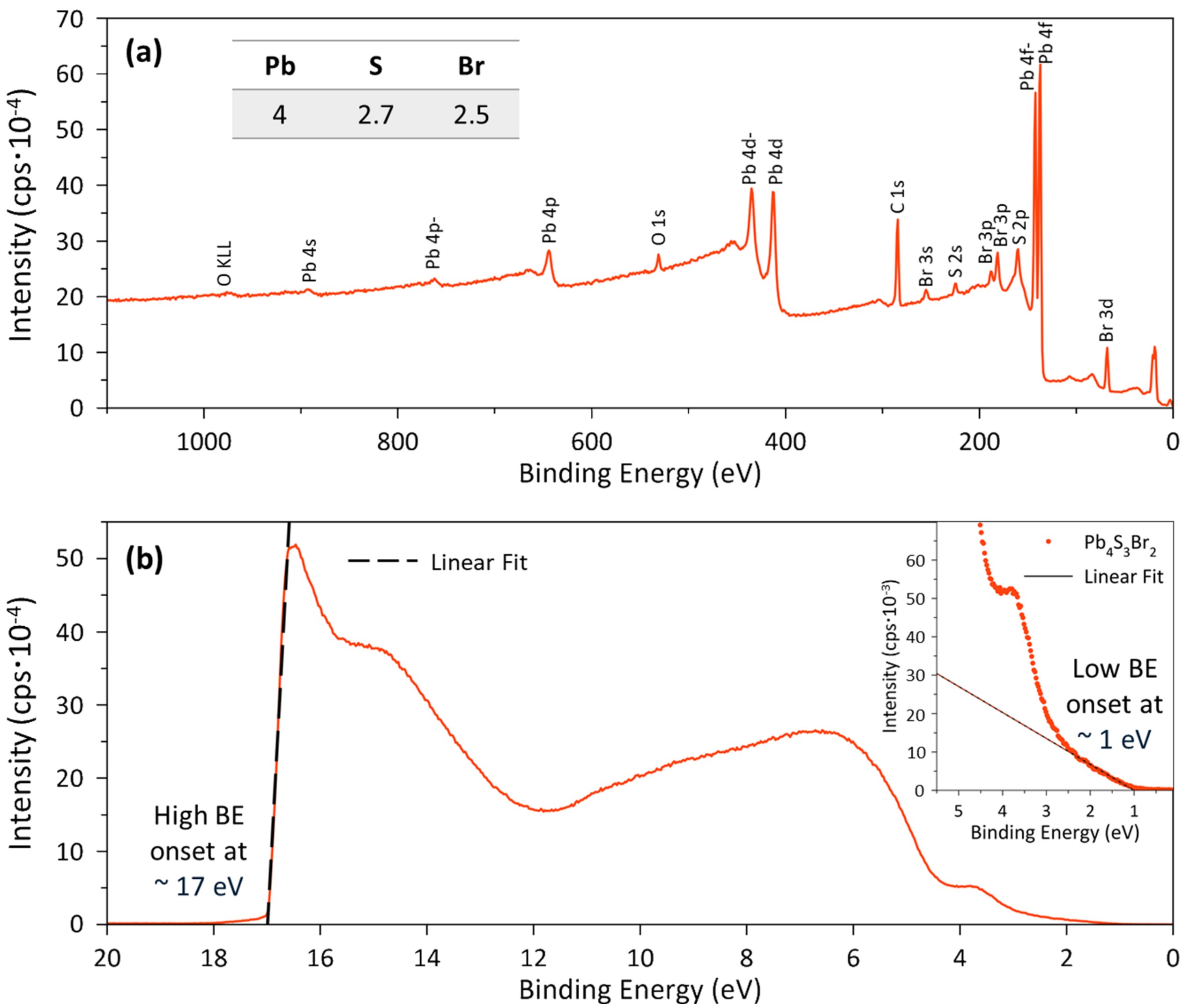

**Figure S11. X-ray Photoelectron Spectroscopy (XPS) and Ultraviolet Photoelectron Spectroscopy (UPS).** a) XPS data for $Pb_4S_3Br_2$. Inset: composition of element by XPS analysis, in agreement with the stoichiometry of the material. b) UPS data for the valence band of $Pb_4S_3Br_2$ that is found at 1.0 eV ± 0.1 eV eV *vs* Fermi Level, since the work function is 4.3 eV, the valence band vs vacuum level is at 5.23 eV. The high and low binding energy onsets were obtained *via* linear extrapolation and are marked.

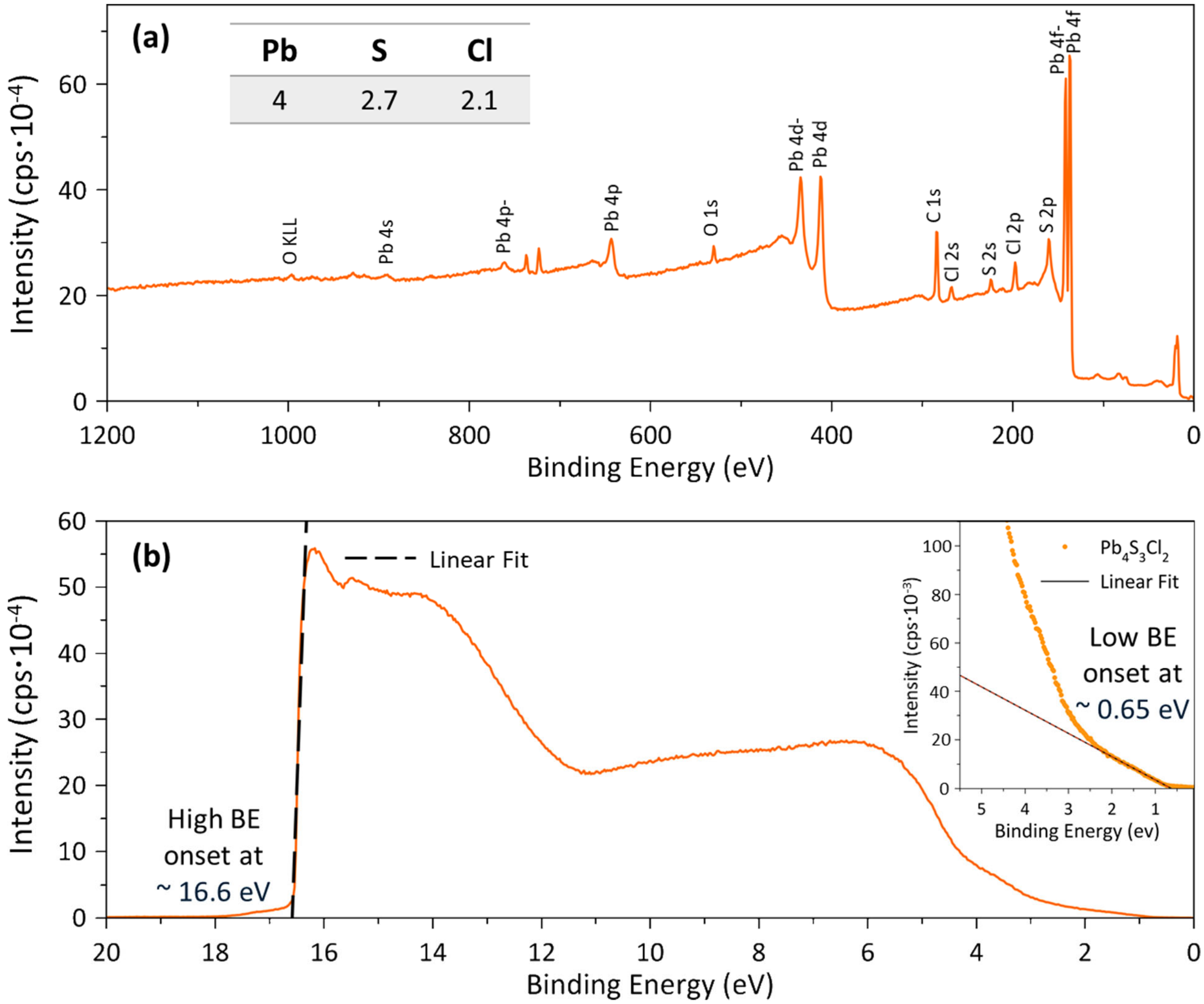

**Figure S12. X-ray Photoelectron Spectroscopy (XPS) and Ultraviolet photoelectron Spectroscopy (UPS).** a) XPS data for $Pb_4S_3Cl_2$. Inset: composition of elements by XPS analysis, in agreement with the stoichiometry of the material. b) UPS data for the valence band of $Pb_4S_3Cl_2$ that is found at 0.65 eV vs Fermi Level, since the work function is 4.67, eV the valence band vs vacuum level is at 5.32 eV. The high and low binding energy onsets were obtained *via* linear extrapolation and are marked.

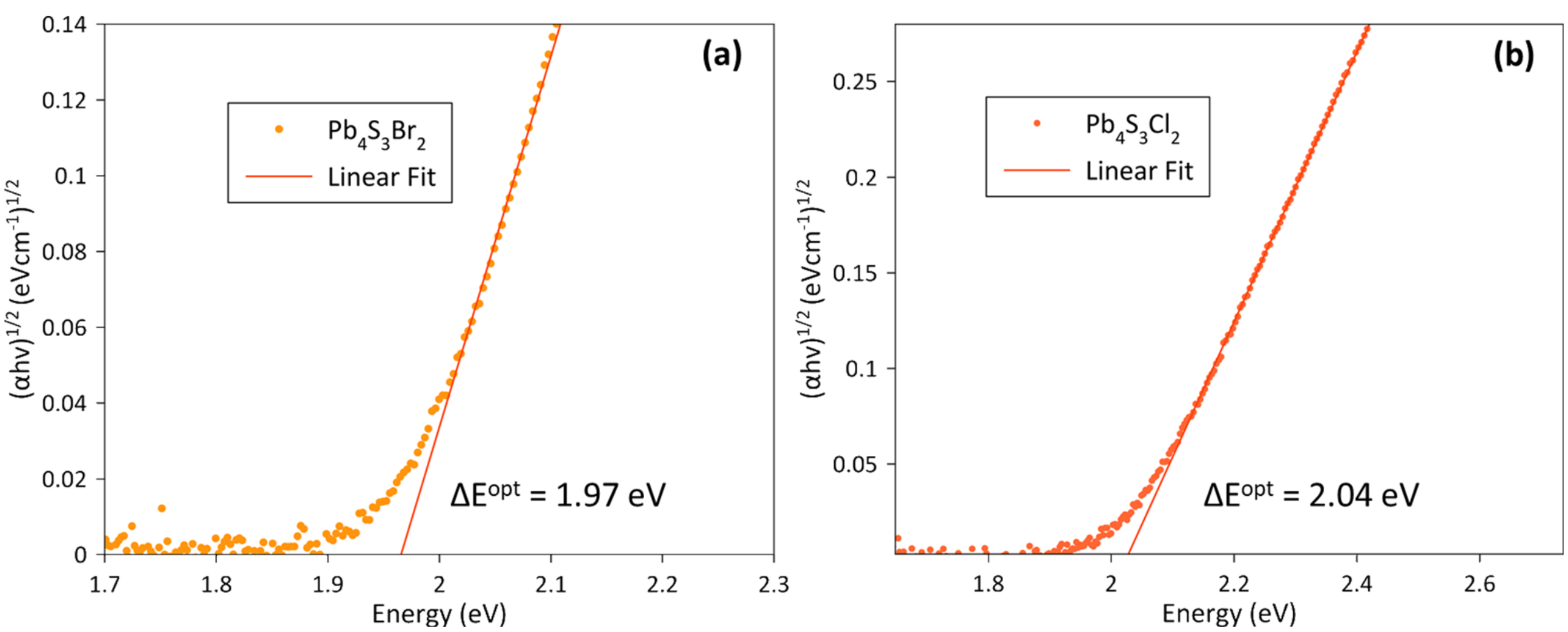

**Figure S13. Tauc plots.** a) Tauc plot from the absorbance spectrum of $Pb_4S_3Br_2$, used to determine the optical band gap of the material. b) Tauc plot from the absorbance spectrum of $Pb_4S_3Cl_2$, used to determine the optical band gap of the material.

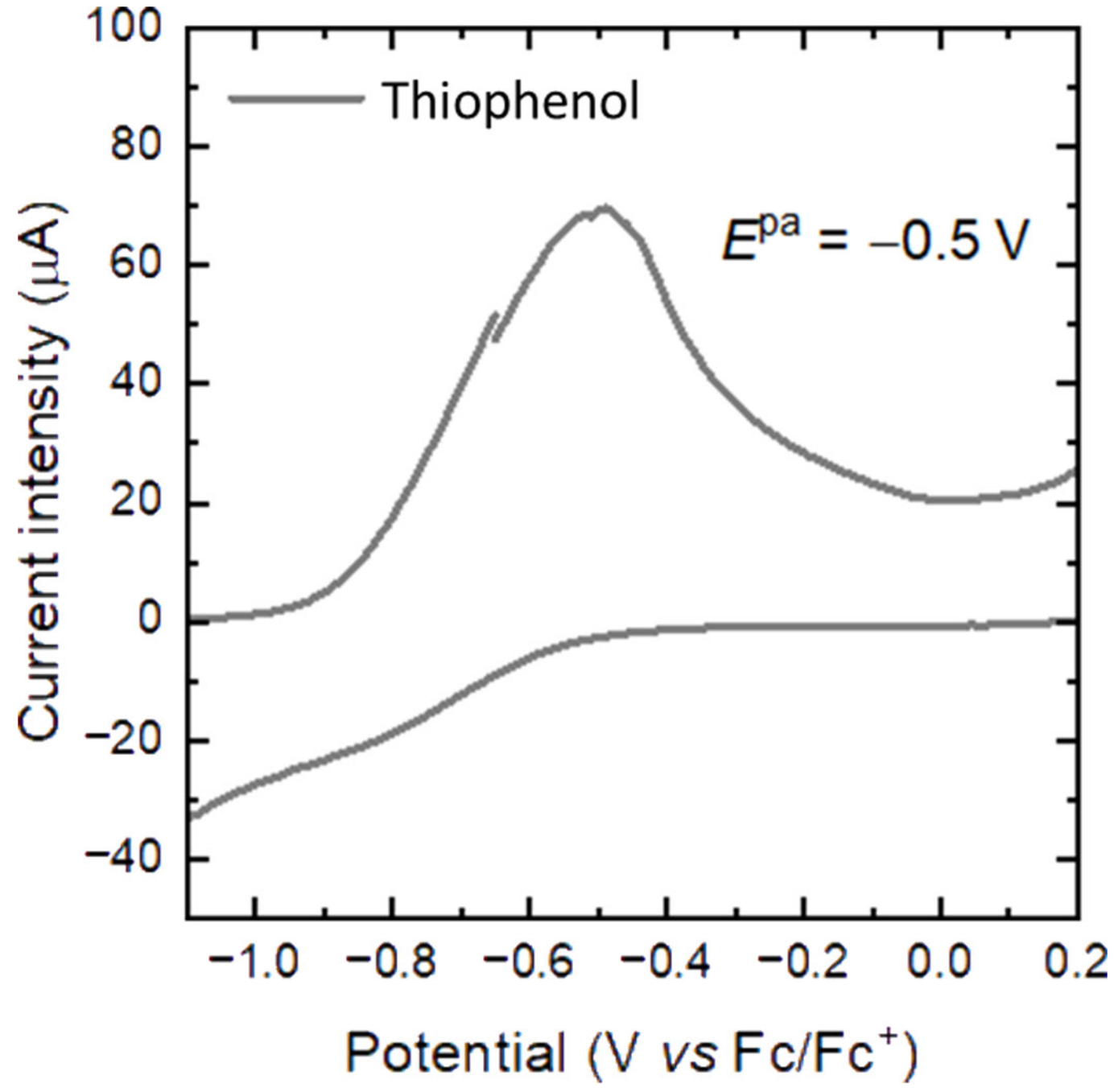

**Figure S14. Electrochemical measurements.** CV curve of thiophenol, used to determine the highest occupied molecular orbital (HOMO) of the molecule (-4.2 eV) at the anodic peak maximum.

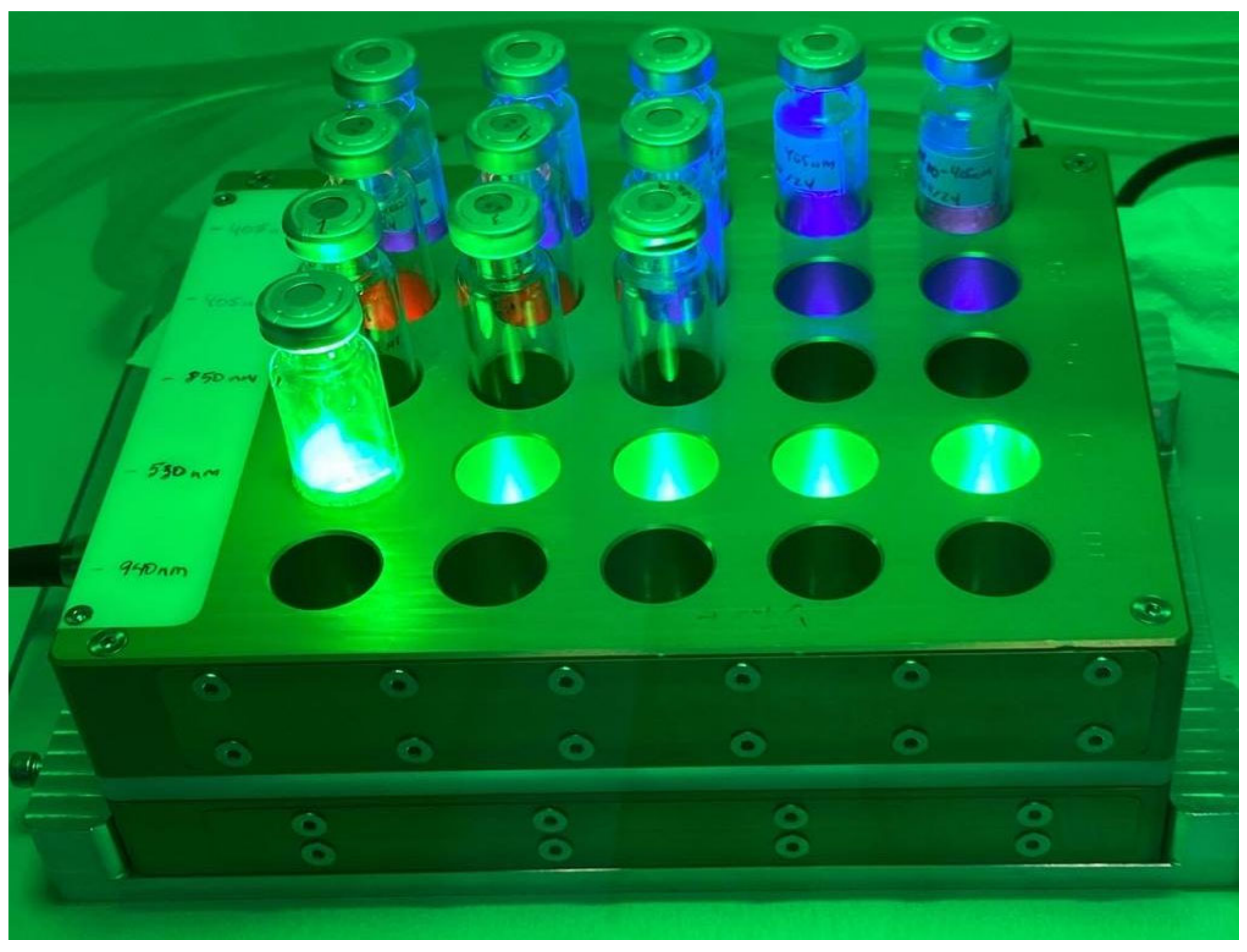

**Figure S15.** The photoreactor used for the photoreactions is equipped with five LED lines, with different wavelength emissions.

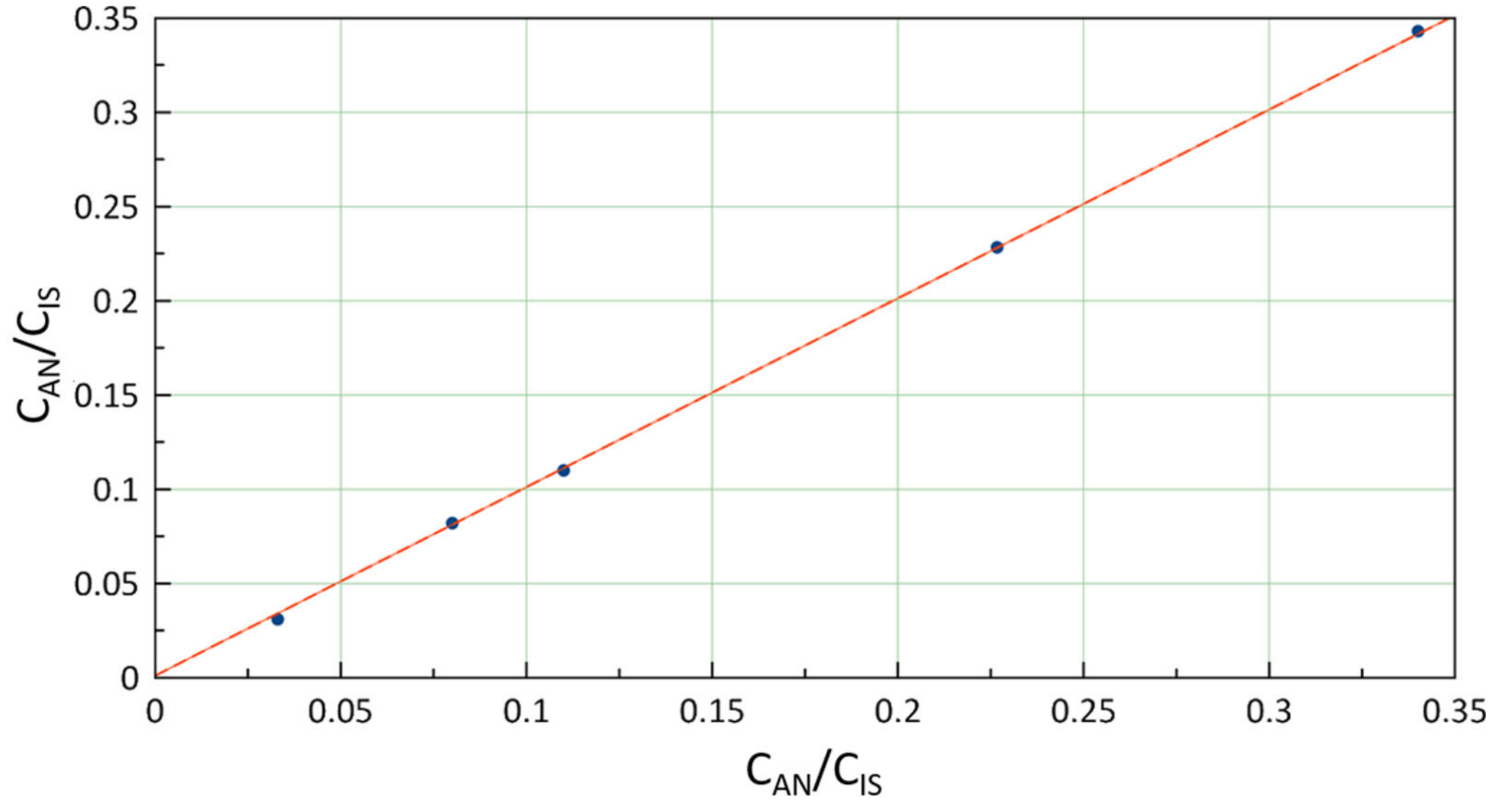

**Figure S16. Calibration curve.** Correlation between concentration and peak area ratio (analyte/IS).

**Table S1.** GC retention time of the different compounds.

| Retention time (min) | Compound |
|---|---|
| 15.122 – 15.128 | Biphenyl (internal standard) |
| 7.627 | Thiophenol |
| 20.510 | Disulfide diphenyl |
| 26.045 | Bis(4-bromophenyl) disulfide |
| 25.454 | Bis(4-methoxyphenyl) disulfide |
| 14.015-14.021 | Cyclohexanone |
| 19.880-19.887 | 1-octadecene |

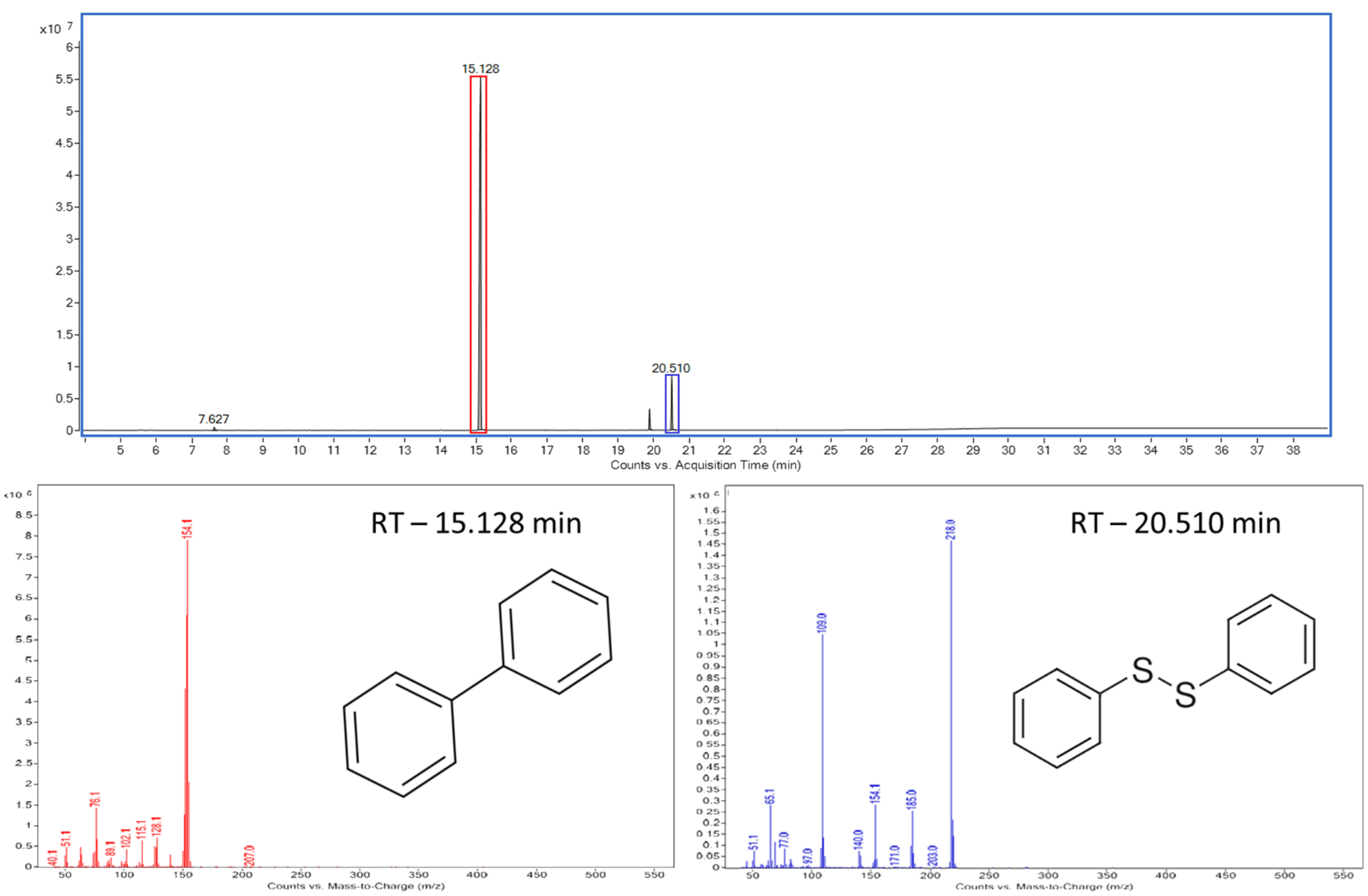

**Figure S17. Gas Chromatography – Mass Spectrometry.** Gas chromatography of the thiophenol reaction in cyclohexane and mass spectra of the internal standard (bottom left) and the product (bottom right).

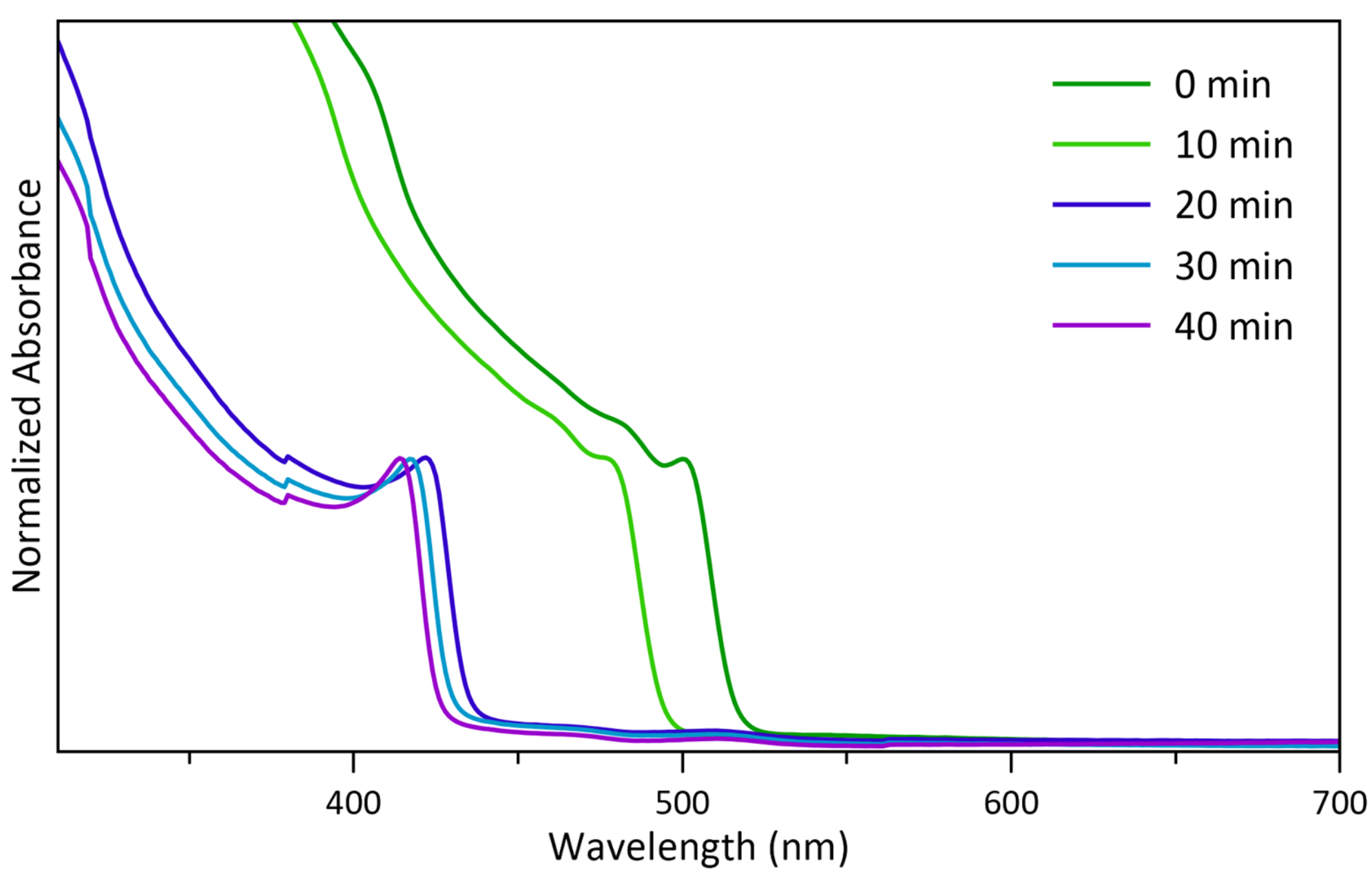

**Figure S18. Absorbance spectra of $CsPbBr_3$ in $CH_2Cl_2$ under light.** Time evolution of the absorption spectrum of $CsPbBr_3$ in $CH_2Cl_2$ under illumination at 450 nm.

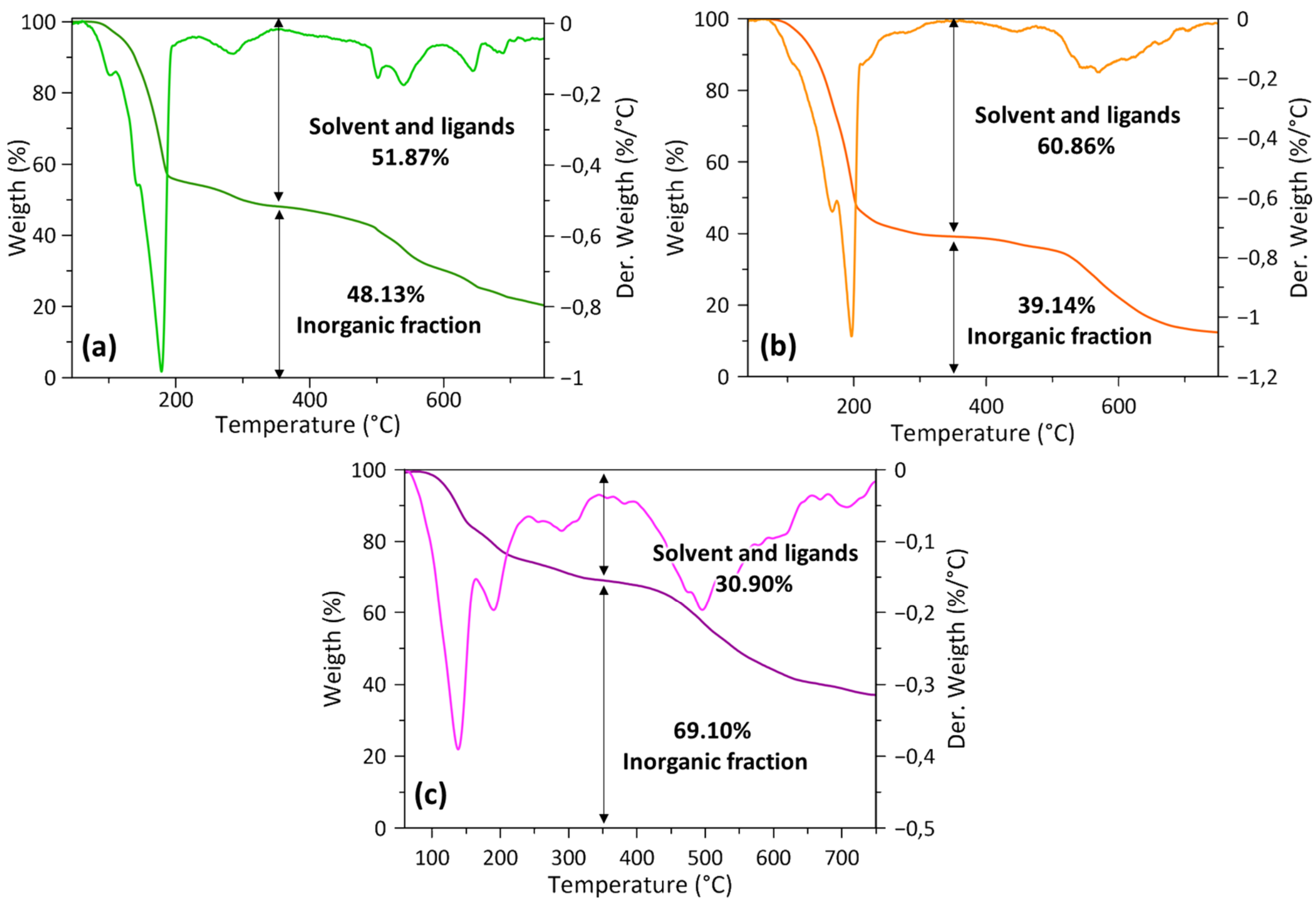

**Figure S19. Thermogravimetric analysis.** a) TGA analysis of $CsPbBr_3$ NCs (green line) and first derivative (light green line). b) TGA analysis of $CsPbBr_3/Pb_4S_3Br_2$ HSs (orange line) and first derivative (light orange line). c) TGA analysis of $Pb_4S_3Br_2$ NCs (purple line) and first derivative (light purple line). Thermogravimetry analyses (TGA) indicated that the first weight loss is due to two different contributions: hexane adsorbed on the NCs surface, as reported previously,[4] and organic ligands (both bound and free ligands). Therefore, the inorganic NC cores accounted for the remaining 39.1% of weight in the case of the heterostructures, 48.1 % in the case of perovskites, and 69.1 % for the chalcohalides.

**Table S2**. Comparison of the photocatalytic performances of $CsPbBr_3/Pb_4S_3Br_2$ HSs and other semiconductor photocatalysts for the oxidative coupling of thiophenol under visible light irradiation.

| Photocatalyst | Substrate | Product yield (%) | Selectivity (%) | Reaction time | Wavelength | TON | TOF | REF |
|---|---|---|---|---|---|---|---|---|
| $CsPbBr_3/Pb_4S_3Br_2$ | Thiophenol | 81 | 87 | 90 min | 450 nm | 14300 | 9560 | This work |
| $CsPbBr_3$ | Thiophenol | 98 | not reported | 6h | White light | - | - | 5 |
| CdS | *p*-$OCH_3$-PhS | 97 | 99 | 4h | White light | - | - | 6 |
| GR–CdS–Co-Pi composite | *p*-$OCH_3$-PhS | 23 μmol | >99 | 5h | White light | - | - | 7 |
| CdSe/CdS | Thiophenol | 87 | not reported | 2h | 415 nm | - | - | 8 |

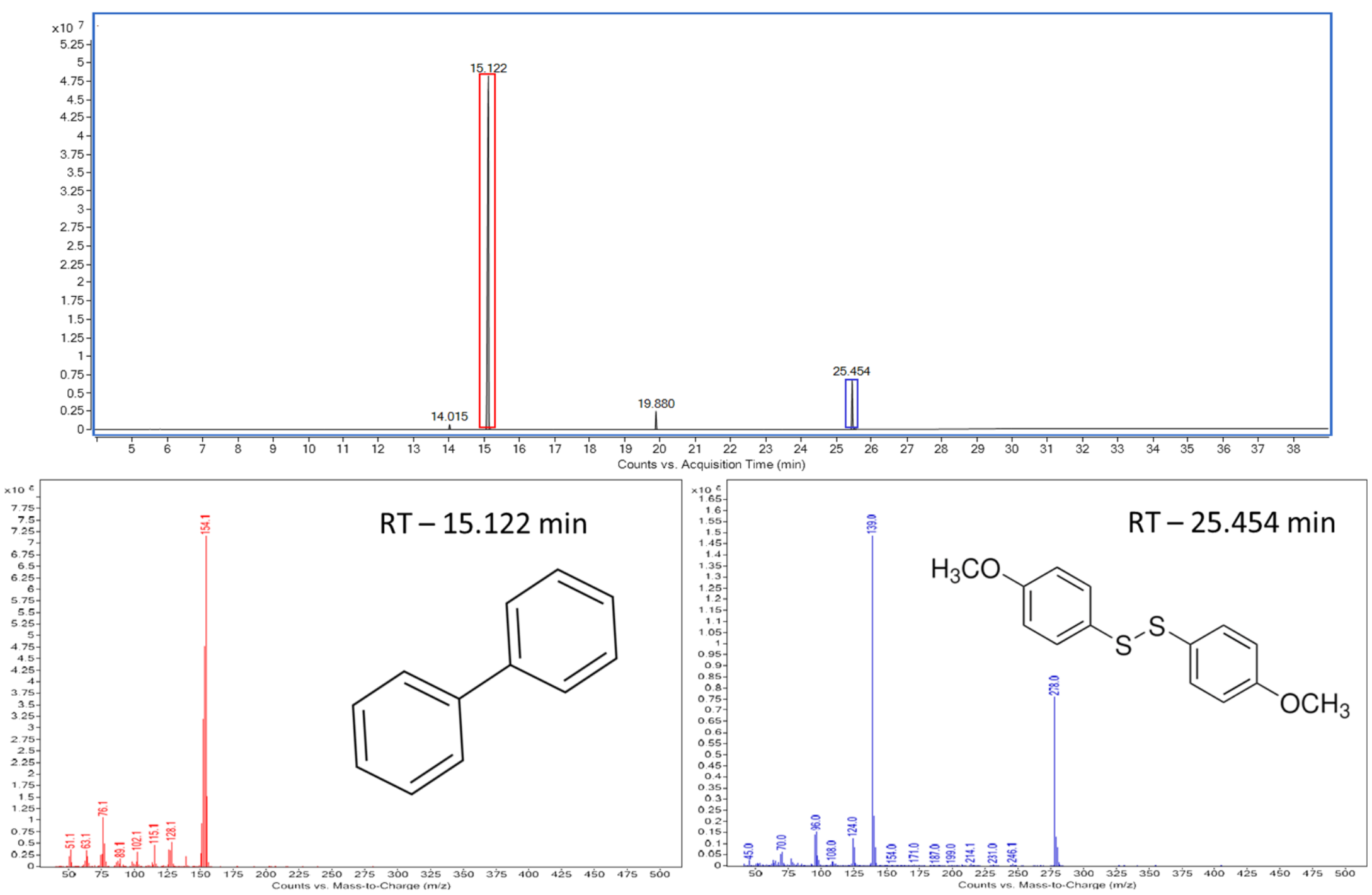

**Figure S20. Gas Chromatography – Mass Spectrometry.** Gas chromatography of the 4-methoxythiophenol reaction in cyclohexane and mass spectra of the internal standard (bottom left) and the product (bottom right).

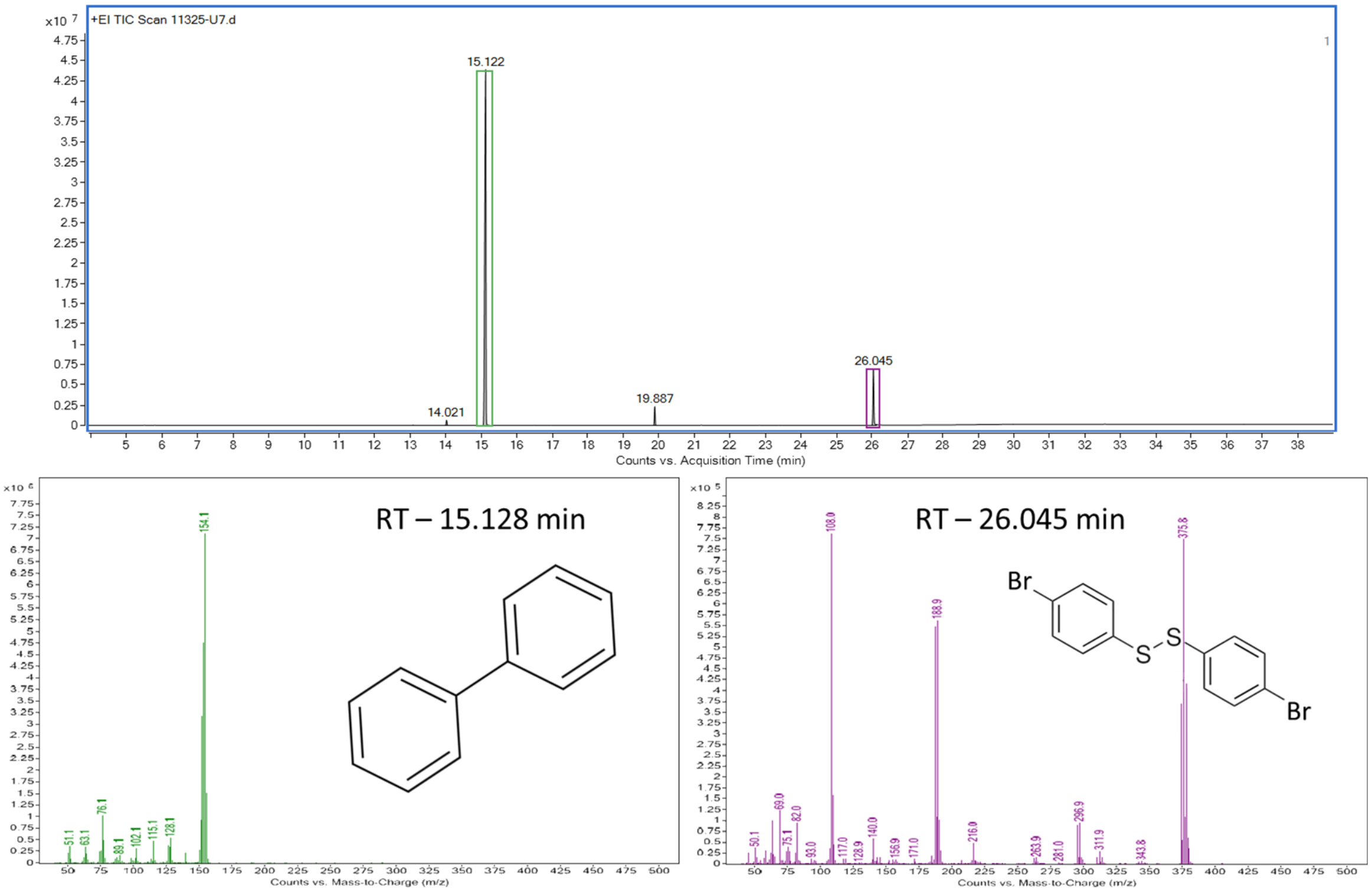

**Figure S21. Gas Chromatography – Mass Spectrometry.** Gas chromatography of the 4-bromothiophenol reaction in cyclohexane and mass spectra of the internal standard (bottom left) and the product (bottom right).

**Table S3.** Standard conditions using different scavengers and $CsPbBr_3/Pb_4S_3Br_2$ HS as photocatalyst for the coupling of the thiophenol.

| Entry | Scavenger | Variance | Conversion (%) | Product 1b (%) |
|---|---|---|---|---|
| **1** | 1,4 Benzoquinone | None | 97 | 23 |
| **2** | DIPEA | None | 99 | 61 |
| **3** | TEMPO | None | 100 | 3 |
| **4** | TEMPO | $N_2$ | 100 | 3 |

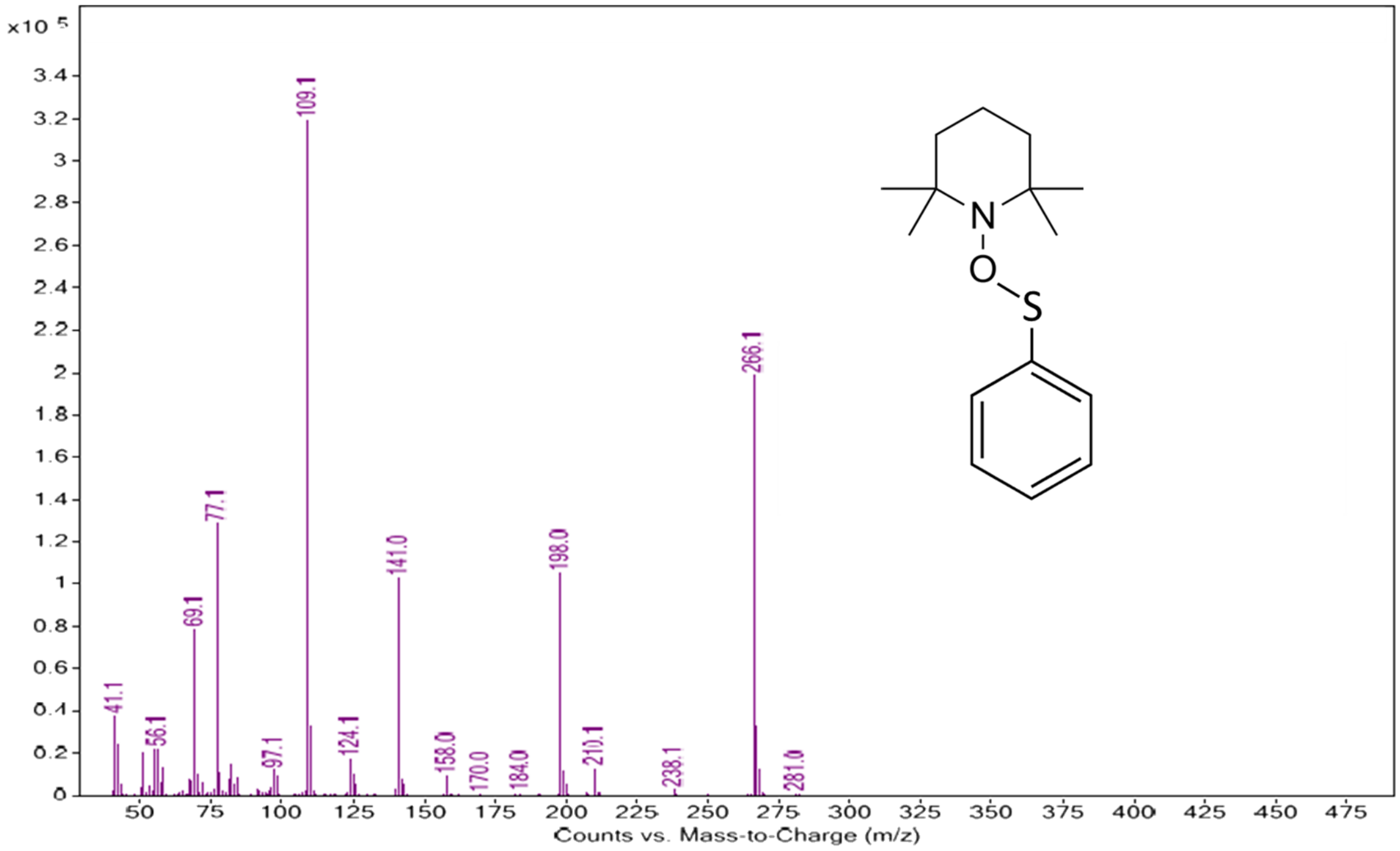

**Figure S22. Gas Chromatography – Mass Spectrometry.** Mass spectrum of the adduct (TEMPO-PhS) between TEMPO and thiyl radical, showing the typical fragments of both units.

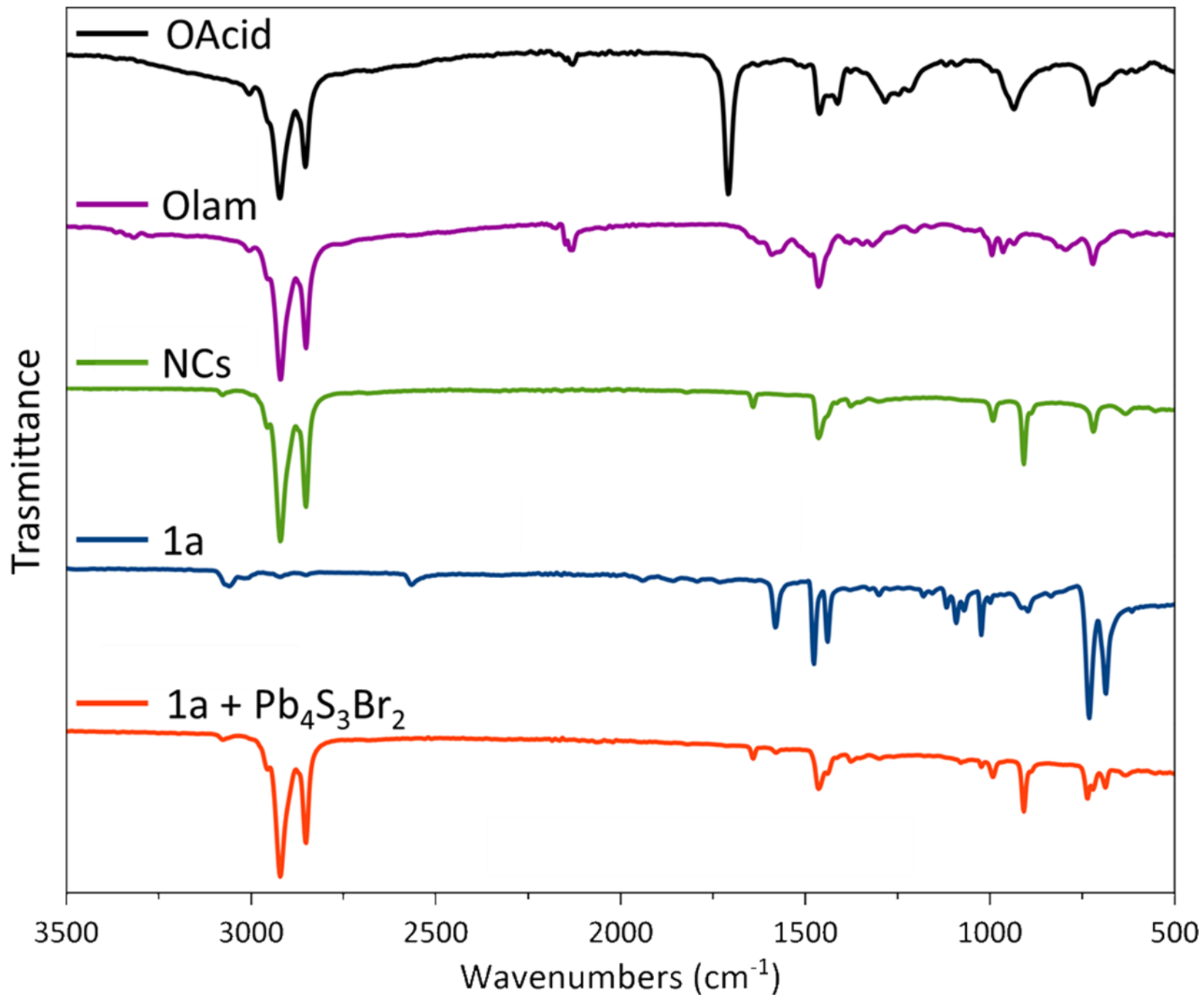

**Figure S23. Fourier Transform Infrared Spectroscopy.** FTIR spectra of OAcid (black line), Olam (purple line), $Pb_4S_3Br_2$ NCs (green line), 1a-thiophenol (blue line) and a mixture of $Pb_4S_3Br_2$ NCs and 1a (red line).

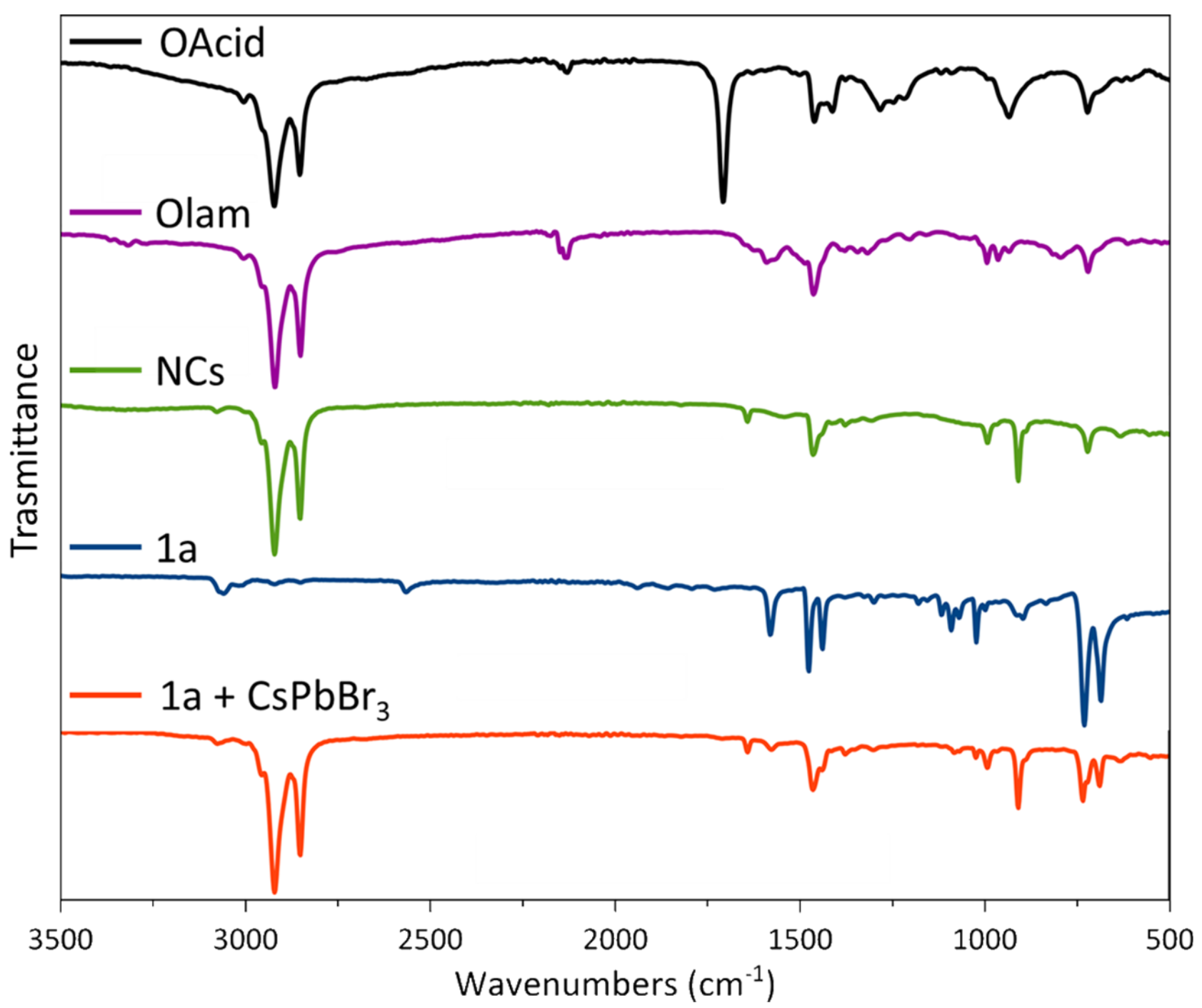

**Figure S24. Fourier Transform Infrared Spectroscopy.** FTIR spectra of OAcid (black line), Olam (purple line), $CsPbBr_3$ NCs (green line), 1a (thiophenol, blue line) and a mixture of $CsPbBr_3$ NCs and 1a (red line).

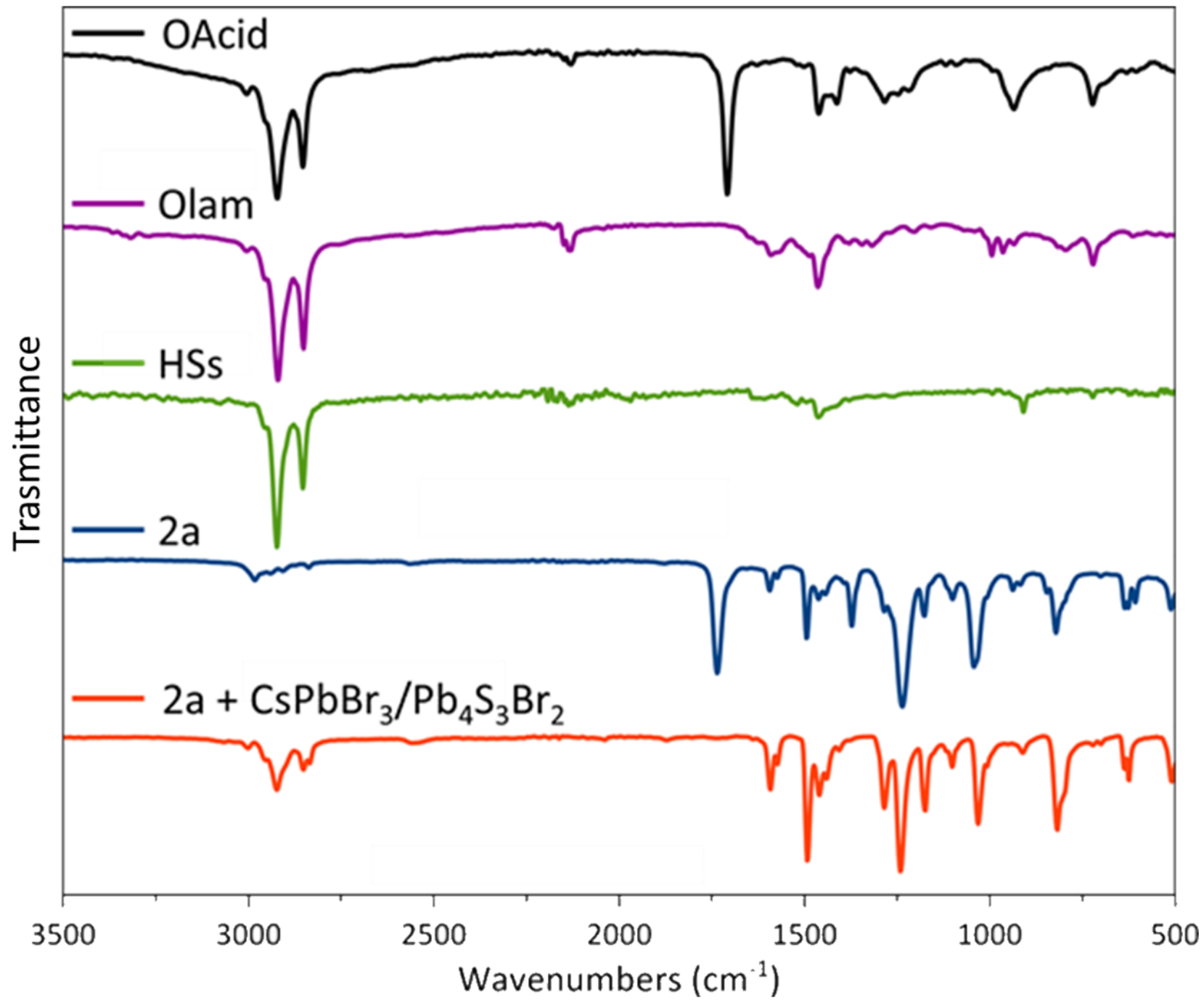

**Figure S25. Fourier Transform Infrared Spectroscopy.** FTIR spectra of OAcid (black line), Olam (purple line), $CsPbBr_3/Pb_4S_3Br_2$ HSs (green line), 2a (4-methoxythiophenol, blue line) and a mixture of $CsPbBr_3/Pb_4S_3Br_2$ HSs and 2a (red line).

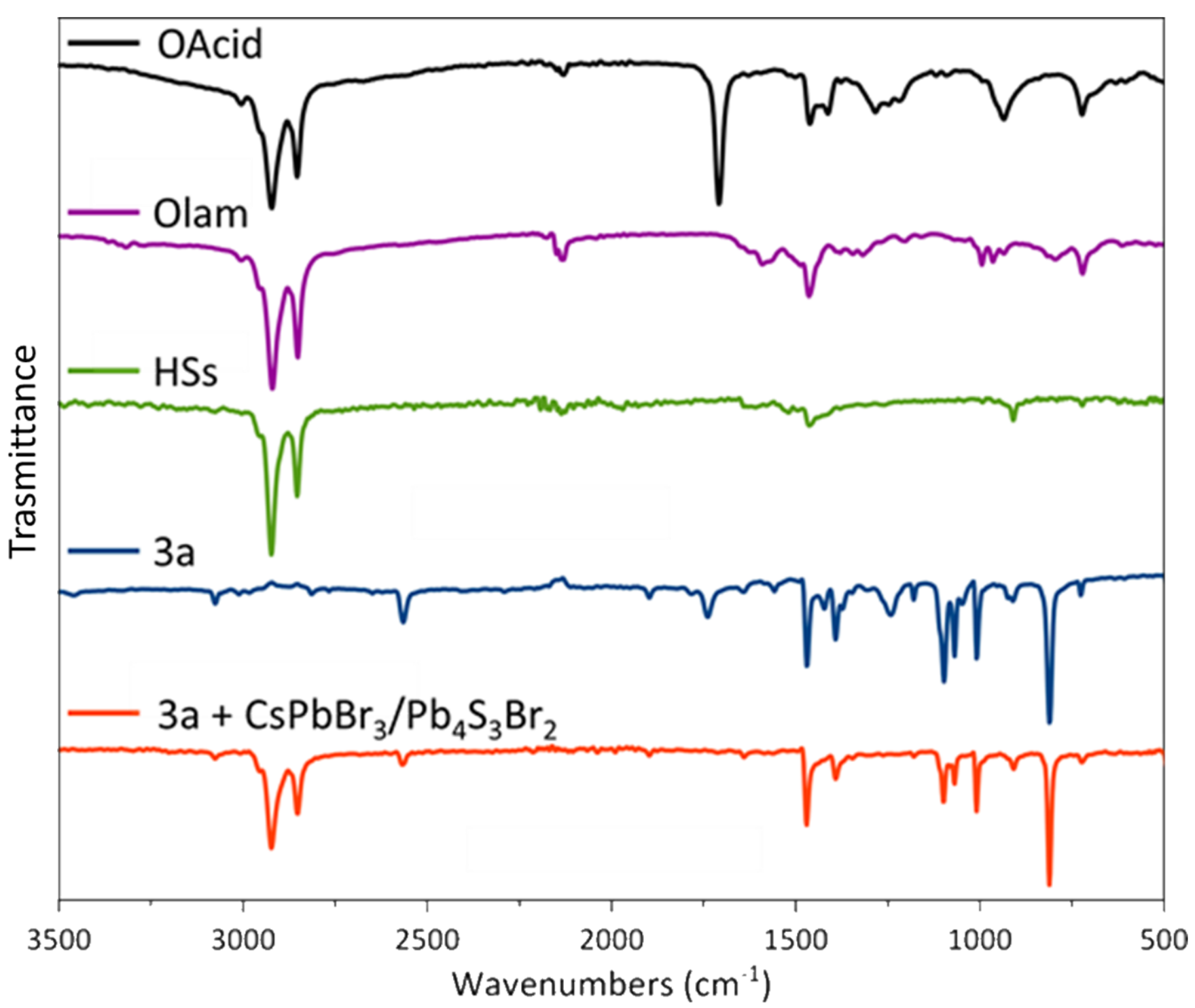

**Figure S26. Fourier Transform Infrared Spectroscopy.** FTIR spectra of OAcid (black line), Olam (purple line), $CsPbBr_3/Pb_4S_3Br_2$ HSs (green line), 3a (4-bromothiophenol, blue line) and a mixture of $CsPbBr_3/Pb_4S_3Br_2$ HSs and 3a (red line).

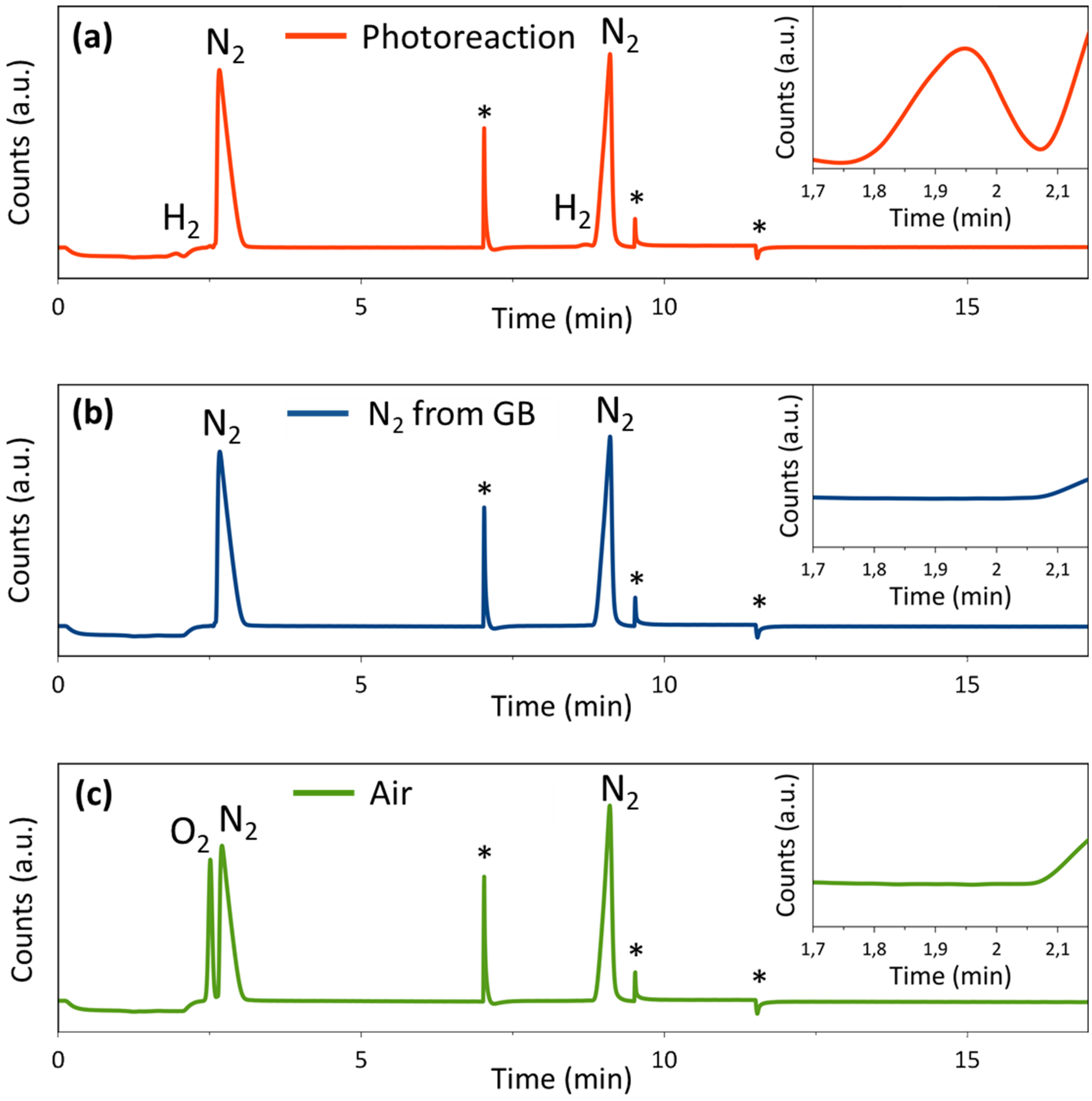

**Figure S27. Typical chromatographs (Thermal Conductivity Detector, TCD) obtained upon the injection of controls and photoreaction samples.** a) TCD trace of the headspace gas from the photoreaction performed under inert atmosphere ($N_2$). Inset: magnification of the region corresponding to the first $H_2$ peak. b) TCD trace of a gas sample collected from the glovebox, showing only $N_2$. Inset: magnification of the $H_2$ region, confirming the absence of hydrogen. c) TCD trace of a gas sample taken from air, showing the presence of $N_2$ and $O_2$. Inset: magnified view of the $H_2$ region, again showing no hydrogen signal. Asterisks (*) identify valve events.

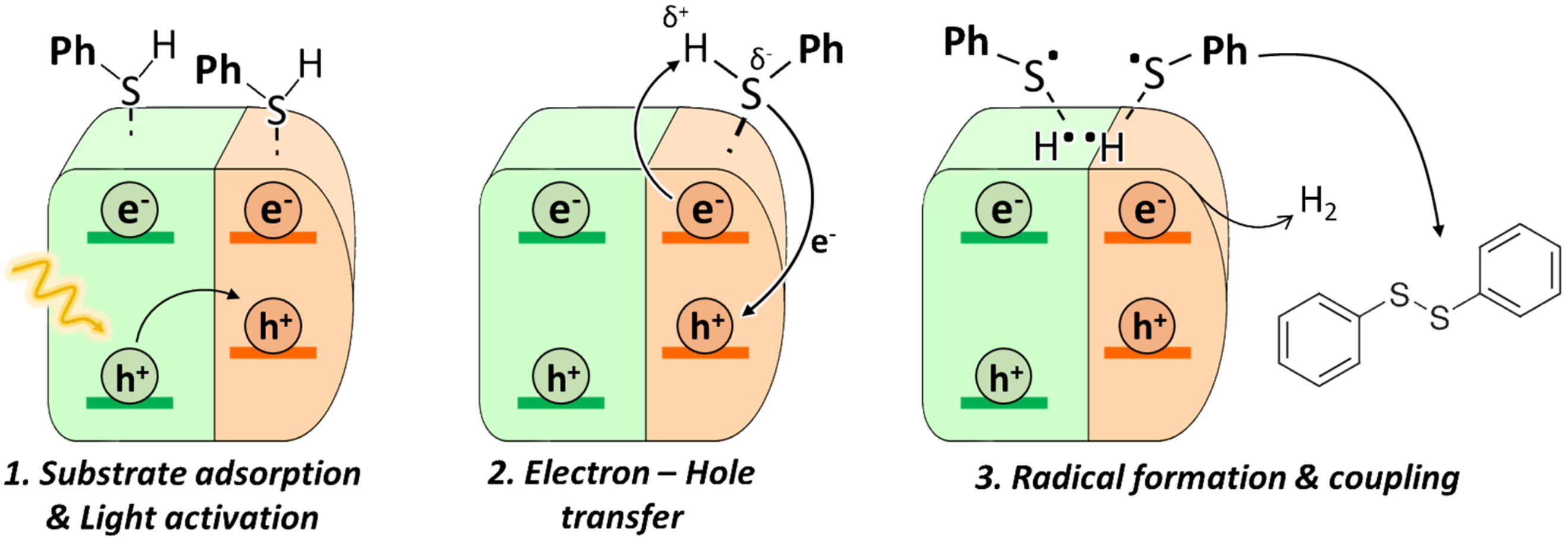

**Figure S28. Plausible reaction mechanism in anaerobic conditions.** Proposed photocatalytic mechanisms for the oxidative coupling of thiophenol under inert atmosphere, with hydrogen and phenyl disulfide formation.

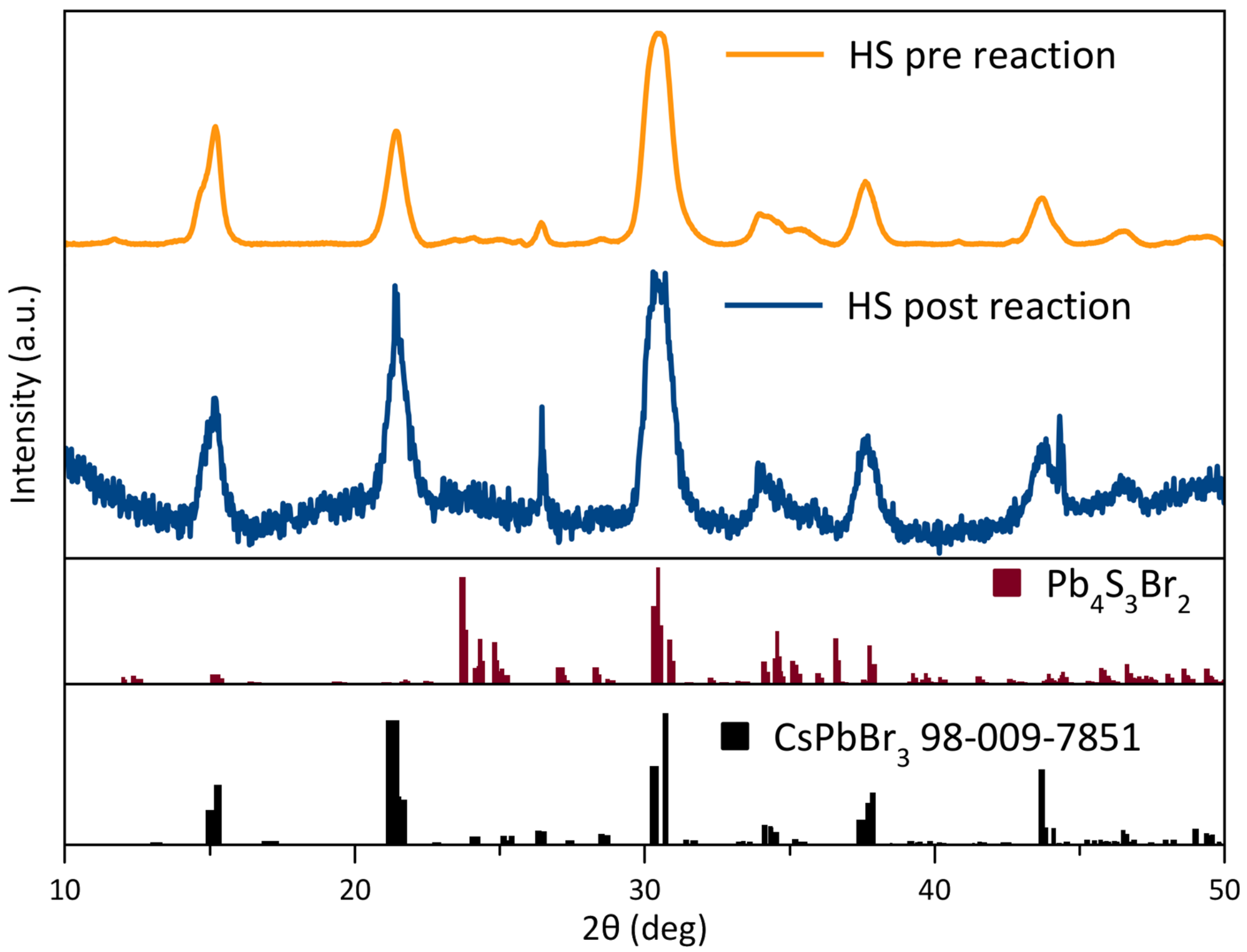

**Figure S29. XRD patterns after photoreaction.** XRD pattern of $CsPbBr_3/Pb_4S_3Br_2$ HSs before (orange line) and after (blue line) photoreaction and $CsPbBr_3$ (black line), $Pb_4S_3Br_2$ (Bordeaux line) as reference.

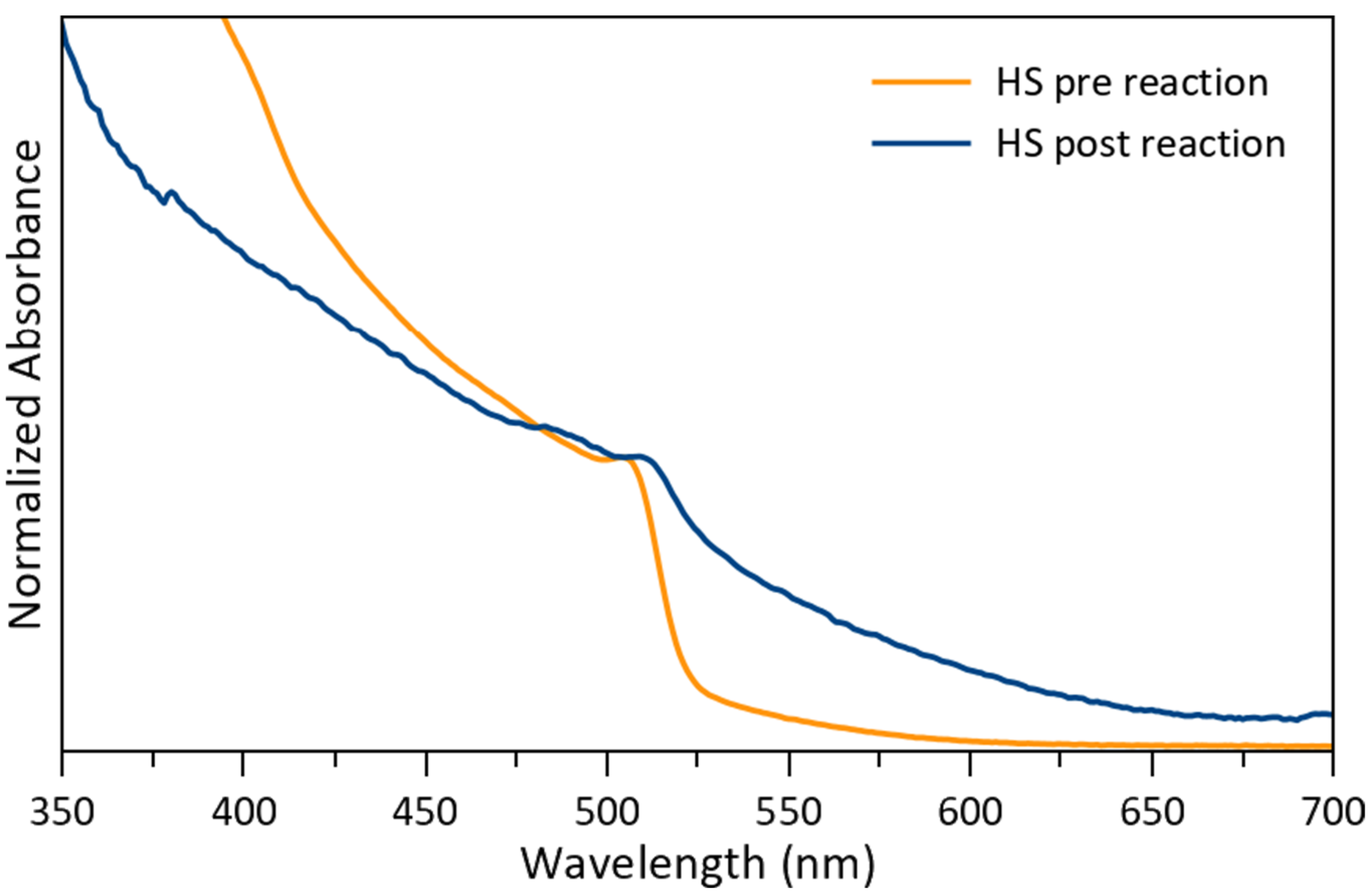

**Figure S30. Absorbance spectra after photoreaction.** Absorbance spectra of $CsPbBr_3/Pb_4S_3Br_2$ HSs before (orange trace) and after (blue trace) photoreaction.

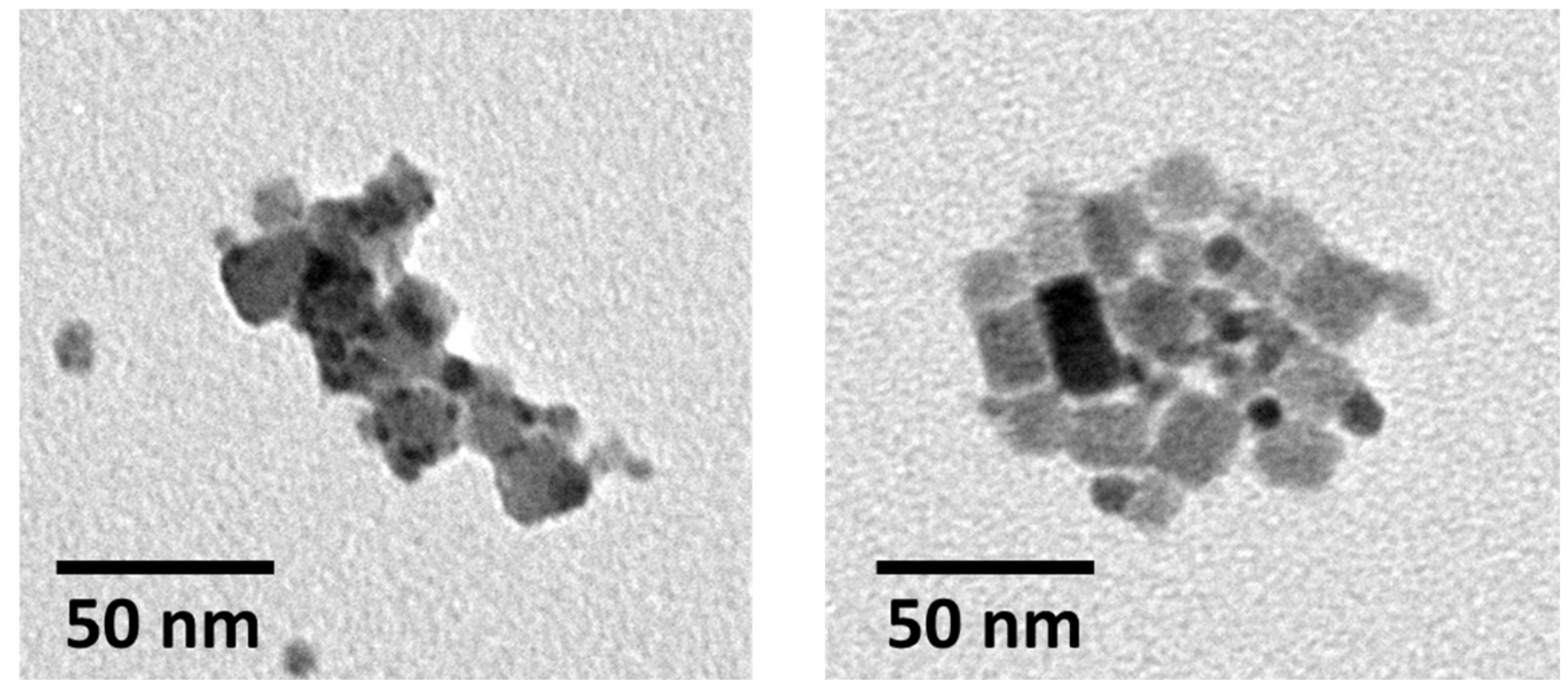

**Figure S31. TEM images after photoreaction.** TEM images of $CsPbBr_3/Pb_4S_3Br_2$ HSs after photoreaction.

**Table S4.** Standard conditions using different photocatalysts (Cl-based) for the coupling of the thiophenol and its derivates.

| Entry | R substituent | Photocatalyst | Conversion (%) | Product (%) | Selectivity (%) |
|---|---|---|---|---|---|
| **1** | -H | $CsPbCl_3$ /$Pb_4S_3Cl_2$ | 99 | 70 | 70 |
| **2** | -H | $CsPbCl_3$ | 99 | 47 | 47 |
| **3** | -H | $Pb_4S_3Cl_2$ | 73 | 44 | 60 |
| **4** | $-OCH_3$ | $CsPbCl_3$ /$Pb_4S_3Cl_2$ | 99 | 86 | 86 |
| **5** | $-OCH_3$ | $CsPbCl_3$ | 76 | 54 | 71 |
| **6** | $-OCH_3$ | $Pb_4S_3Cl_2$ | 80 | 52 | 65 |
| **7** | -Br | $CsPbCl_3$ /$Pb_4S_3Cl_2$ | 95 | 77 | 81 |
| **8** | -Br | $CsPbCl_3$ | 87 | 62 | 71 |
| **9** | -Br | $Pb_4S_3Cl_2$ | 71 | 27 | 38 |

**Table S5.** Standard condition using different photocatalysts (I-based) for the coupling of the thiophenol and its derivates.

| Entry | R substituent | Photocatalyst | Conversion (%) | Product (%) | Selectivity (%) |
|---|---|---|---|---|---|
| **1** | -H | $CsPbI_3$/$Pb_4S_3Br_2$ | 100 | 80 | 80 |
| **2** | -H | $CsPbI_3$ | 86 | 52 | 61 |
| **3** | $-OCH_3$ | $CsPbI_3$/$Pb_4S_3Br_2$ | 100 | 83 | 83 |
| **4** | $-OCH_3$ | $CsPbI_3$ | 100 | 82 | 82 |
| **5** | -Br | $CsPbI_3$/$Pb_4S_3Br_2$ | 100 | 64 | 64 |
| **6** | -Br | $CsPbI_3$ | 100 | 60 | 60 |